\documentclass[12pt]{article}

\usepackage{scicite}
\usepackage{times}
\usepackage{graphicx,caption}
\usepackage{amsmath,amsthm,bm}
\usepackage{amssymb}  
\usepackage{color}
\usepackage{colortbl}
\usepackage{mathtools}
\usepackage{relsize}
\usepackage{mdwlist}
\usepackage{url}
\usepackage[displaymath,mathlines]{lineno}
\usepackage[labelformat=empty]{caption}
\usepackage[normalem]{ulem}
\usepackage{wasysym}
\usepackage{comment}

\newenvironment{sciabstract}{%
\begin{quote} \bf}
{\end{quote}}

\title{Tactile Displays Driven by Projected Light}

\author
{Max Linnander$^{1}$,  Dustin Goetz$^1$,  Gregory Reardon$^3$, \\ Vijay Kumar$^1$, Elliot Hawkes$^1$, and Yon Visell$^{1,2,3,4*}$\\
\\
\normalsize{$^1$Department of Mechanical Engineering, $^2$Department of Bioengineering,}\\ 
\normalsize{$^3$Media Arts \& Technology Program,} 
\normalsize{$^4$Department of Electrical and Computer Engineering,}\\
\normalsize{University of California, Santa Barbara, USA.}\\
\normalsize{$^\ast$Corresponding author. E-mail:  yonvisell@ucsb.edu.} }
\date{}

\begin{document}

\baselineskip24pt

\maketitle 

\begin{sciabstract}
Tactile displays that lend tangible form to digital content could transform computing interactions.  However, achieving the resolution, speed, and dynamic range needed for perceptual fidelity remains challenging.  We present a tactile display that directly converts projected light into visible tactile patterns via a photomechanical surface populated with millimeter-scale optotactile pixels.  The pixels transduce incident light into mechanical displacements through photostimulated thermal gas expansion, yielding millimeter scale displacements with response times of 2 to 100 milliseconds.  Employing projected light for power transmission and addressing renders these displays highly scalable.  We demonstrate optically driven displays with up to 1,511 addressable pixels -- several times more pixels than any prior tactile display attaining comparable performance. Perceptual studies confirm that these displays can reproduce diverse spatiotemporal tactile patterns with high fidelity. This research establishes a foundation for practical, versatile high-resolution tactile displays driven by light.
\end{sciabstract}

\section*{One-Sentence Summary}

Photomechanical surfaces directly convert projected light into visible and tactile patterns.




\section*{Introduction}

Tactile displays give physical form to digital content, conveying information through tactile patterns that can represent data, user interfaces, or virtual objects. They could profoundly transform human-computer interaction, paralleling the role of visual displays in driving successive revolutions in computing, from the first widespread computers of the 1970s, to the mobile, wearable, and virtual reality systems of today \cite{van1966computer}.  Applications of tactile displays span many fields, including automotive interfaces that emulate physical controls with software reconfigurability \cite{colgate2013haptic,serafin2007user,ng2017evaluation}, educational tools such as animated haptic books \cite{watanabe2006practical,cingel2015augmenting}, and user interfaces for virtual and augmented reality \cite{wang2019multimodal,benko2016normaltouch}.  However, creating multipurpose tactile displays with perceptual fidelity imposes stringent requirements on spatial resolution ($\lesssim$ 3 mm), response time ($<$50 ms), displacement ($>$ 0.2 mm), and force ($>$ 5 mN),  with proportional output \cite{kern2023engineering,jones2006human,copeland2010identification,craig1987vibrotactile}.  Achieving these requirements in displays of useful size at sustainable costs will require significant advancements in actuation technologies, scalability, integration density, and fabrication methods.

A variety of materials and actuation methods have been studied for tactile display engineering, including electrostatic, piezoelectric, electromagnetic, fluidic, and electroosmotic methods \cite{leroy2020haxel,grasso2023haxel,schultz2023flatpanel,streque2012emactuator,wood2005piezeo,heisser2021combustion,biswas2019materials}.  However, the state of the art remains dominated by bespoke displays with a hundred or fewer pixels; see Table S1 and survey \cite{biswas2019materials}.\footnote{Braille graphics devices could be cited as exceptions, but are generally designed to represent static graphics or text. They use binary actuation with long refresh times, thus do not meet performance requirements for general purpose tactile display  \cite{yang2021survey}.}   
 A major challenge is scaling electronic tactile displays to larger sizes and resolutions at sustainable cost and complexity while maintaining performance parameters. As size and resolution increase, 
the complexity of driving and control circuits grow rapidly.  Moreover, modifying the configuration of such systems often requires electronic redesign, hindering development.

An intriguing alternative is to exploit light transmission for tactile actuation.   Tactile displays are often designed to integrate with video displays that utilize optical transmission. Harnessing light for tactile actuation presents a synergistic opportunity that could offer practical and performance benefits. Optical energy can be modulated at high frequencies and directed with high spatial precision using mature technologies, including video and laser projection, enabling rapid and localized wireless energy delivery. Even compact laser modules can deliver substantial power via light (e.g., 5 W), as much as would be required to drive dozens of miniature vibration actuators  \cite{vybronics,choi2012vibrotactile}.  Utilizing projected light for tactile displays may simplify power distribution and control, reducing system complexity, and supporting scalability.

Despite its potential, targeted light transmission for tactile actuation has been sparsely investigated. Prior studies have employed thermal mechanisms for photomechanical transduction, by converting absorbed light into heat, which is then transformed into tactile actuation. Examples include solid thermoelastic expansion \cite{hwang2021bimorph}, phase transitions in liquid crystal elastomers \cite{torras2014optomechanical,camargo2012opticalbraille}, 
and liquid vaporization \cite{hiraki2020laserpouch}. Due in part to constraints arising from heat transfer, these systems have often exhibited slow response times of seconds or longer (Table S2).  Employing a captive gas as the working medium can enable tactile displays to achieve the rapid response times that are required for perceptual fidelity.  Gases exhibit high thermal diffusivity, $\alpha$, due to their low density. High diffusivity promotes rapid heat transfer, because the heat transfer timescale is given by $\tau = L^2/\alpha$, where $L$ is a characteristic length. The low density of gases also results in large temperature increases $\Delta T = Q / (c \rho V)$, where $Q$ is heat input, $V$ is volume, and $c$ is specific heat capacity. Larger values of $\Delta T$ result in faster thermal responses because heat is transferred out of the actuator at a rate described by  Newton's law of cooling, $\dot{Q} = h A \Delta T$, where $h$ is a heat transfer coefficient and $A$ is effective surface area. 

Directly heating a captive gas with light is challenging, as gases are generally transparent. However, efficient heat transfer to a gas can be achieved by integrating a low thermal mass photoabsorber within a gas-filled cavity. 
Such a configuration may enable rapid actuation cycles suitable for tactile displays driven by projected light. This principle has historical parallels in experiments by Bell, Tyndall, and R\"ontgen in the 1880s, 
where intermittent illumination of absorbing materials in gas-filled cavities was observed to yield audible sound \cite{bell1880photoacoustics,tyndall1881photoacoustics,roentgen1881photoacoustics}. Similar principles have been exploited in prior research on remote power transmission for aerospace \cite{garbuny1976optimization}, thermodynamic actuation \cite{okamura2008efficiency}, and optically driven micromechanical valves \cite{mckenzie1992high,hockaday1990direct,gurney1984photofluidic}, among other systems.

\begin{figure*}[ht!]
    \centering
    \vspace{-5mm}
    \includegraphics[width = 125 mm]{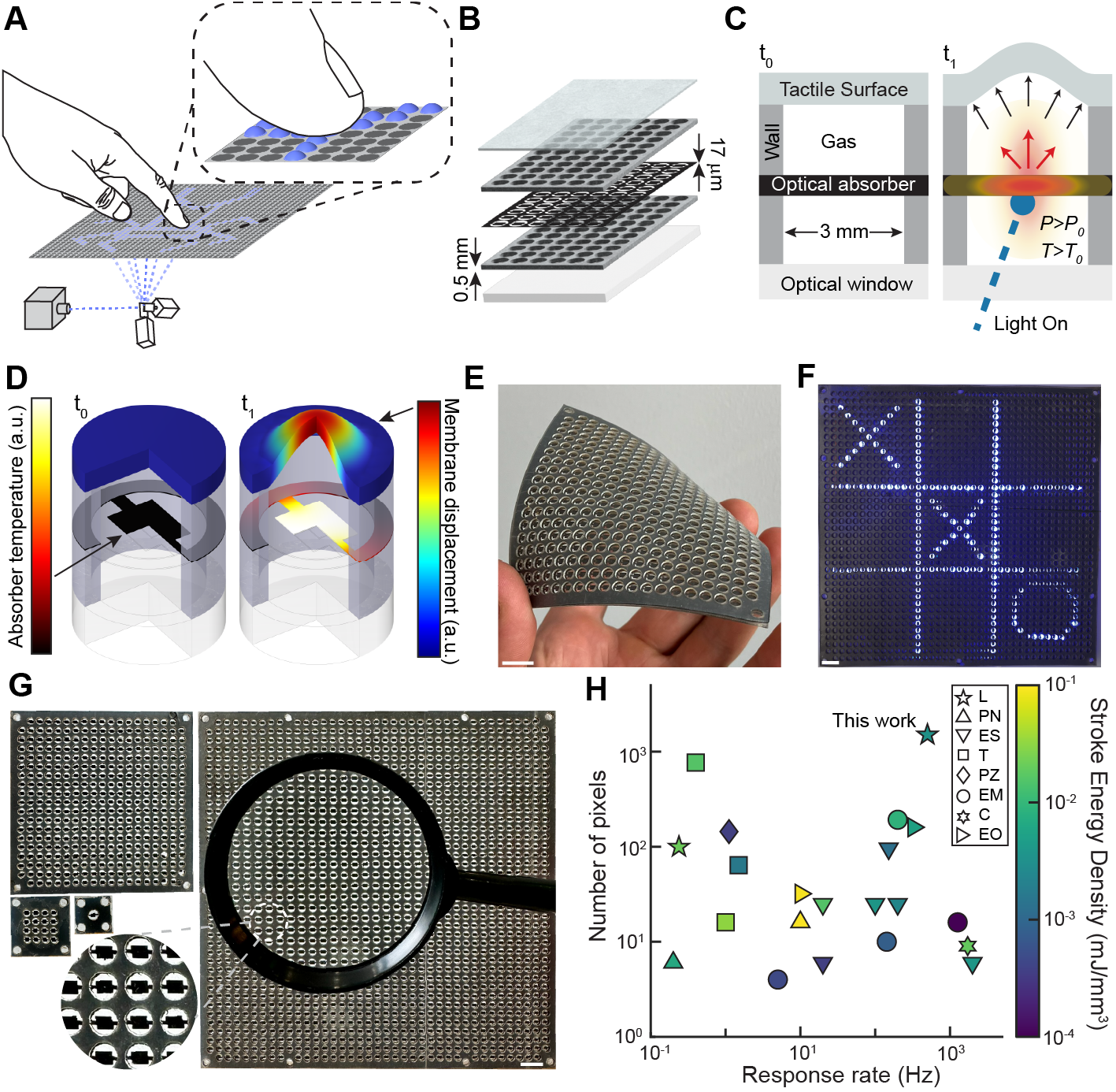}
    \caption{\textbf{Fig. 1. Tactile display driven by projected light.} (\textbf{A}) The display directly converts projected light into visible, tactile patterns via an array of optotactile pixels. (\textbf{B}) Light is converted to forces and displacements via an assembly of patterned layers, supporting manufacturing and scalability. From top: Elastic membrane (EcoFlex 00-10), cavity walls (polysiloxane, PS), thin photoabsorber (PGS, thickness: 17 $\mu$m), cavity walls (PS), optically transparent layer (acrylic). (\textbf{C}) Pixel operation (section view). An incident light pulse is converted into heat by the photoabsorber.  Heat is transferred to air in the cavity, raising gas temperature $T_{gas}$ and pressure $P$.  Gas expansion drives the deflection of the elastic membrane. (\textbf{D}) Finite element analysis of pixel thermo-mechanical response. (\textbf{E}) Flexible tactile display, 357 pixels.  Scale bar: 1 cm. (\textbf{F}) Photostimulation elicits both tactile and optical feedback, due to the transmission of scattered light through the membrane. Extended exposure image of 1,511 pixel display.  Scale bar: 1 cm. (\textbf{G})  This technique can be used to realize displays that vary in dimensions and pixel count. From upper left:  1, 8, 357, and 1,511 pixels. 
     Scale bar: 1 cm. (\textbf{H}) Comparison with prior tactile displays (Fig. S15, Table S1).  Our display enables rapid response rates, scales to pixel counts surpassing prior work, and is uniquely capable of both visual and tactile display. Driving mechanisms: L = light, EM = electromagnetic, PN = pneumatic, ES = electrostatic, T = thermal, PZ = piezoelectric, C = combustion, and EO = electroosmosis. }
    \label{fig:concept}
    \vspace{-5mm}
\end{figure*}

Building upon these insights, we present a tactile display that uses projected light to activate tactile patterns across a surface populated with small optotactile pixels ($L$ = 3 mm). Each pixel contains a thin film photoabsorber suspended in a captive gas. Photostimulation of a pixel drives rapid heating, inducing transient gas expansion (Fig.~\ref{fig:concept}).   Light pulses with energies of 2.5 to 125 mJ are sufficient to activate localized displacements of up to 1 mm and forces of up to 55 mN, with response times of 2 to 100 ms.  The actuation method, optical addressing technique, and simplicity of fabrication make this technique practical and scalable.  We present implementations with up to 1,511 independently addressable pixels distributed over a $15\times15$ cm$^2$ area -- several times more pixels than any previous tactile display approximating the requirements cited above  (Fig.~1H, Table S1).  We show that these light-driven tactile displays are capable of representing a wide variety of spatiotemporal tactile patterns that can be accurately perceived.

\subsection*{Design and Photomechanical Transduction Principle}

The display reproduces visible, tactile patterns by translating projected light into mechanical signals, using an array of optically addressed pixels   (Fig.~\ref{fig:concept}A,B). Each pixel operates on a thermodynamic principle, transducing optical energy into heat, and heat into work. 
Light entering the pixel is converted to heat by a thin photoabsorber suspended in a captive gas (Fig.~\ref{fig:concept}C). The absorber rapidly increases in temperature, transferring heat to the gas, which undergoes a  pressure increase dictated by the ideal gas law $PV = n R T_{gas}$. 
Gas expansion drives the deflection of an elastic membrane sealing the pixels, yielding localized mechanical forces $F(t)$ and displacements $z(t)$ that can be felt by touching the membrane (Fig.~\ref{fig:concept}D).

This general strategy may be used to realize a large variety of displays. The following design choices exemplify one way it can be implemented. A multi-layered architecture was used to facilitate ease of manufacturing and scalability (Fig.~\ref{fig:concept}B). Transparent acrylic forms a base layer, functioning as the optical window (Fig. S2).  An array of cylindrical pixel cavities is formed from two polysiloxane (PS) layers that are patterned via laser cutting  
(Fig.~\ref{fig:concept}C, S4-5).  The pixel dimensions (diameter 3 mm, height 1 mm) were selected based on perceptual requirements,   theoretical analysis, and numerical simulations (Text~S1, S2). Between the pixel cavity layers is a   
pyrolytic graphite sheet (PGS, 17 $\mu$m) that is patterned to form an array of photoabsorbers suspended 
by narrow bridge features (Fig.~\ref{fig:concept}C-D, S4).  PGS was selected due to its high thermal conductivity,  broad spectrum photoabsorption (Fig.~S2), 
and thermal stability\footnote{PGS is stable up to 3000~$^{\circ}$C;  Sustained temperatures over 400$^{\circ}$ C may favor the use of  a more inert gas, such as nitrogen or helium, to inhibit oxidization.}.  Capping this assembly is the tactile surface -- an elastomer membrane 
that seals the air-filled cavities. The display can be made flexible through the use of a thinner or more elastic transparent base layer (Fig.~\ref{fig:concept}E). 

The pixels are stimulated by brief low-energy light pulses (2.5 to 125 mJ) delivered from a compact diode laser  ($\lambda=450$ nm; Fig.~\ref{fig:concept}C, S6). Optical addressing is facilitated by commodity xy  scanning galvanometer mirrors, yielding spatiotemporal tactile patterns that can be felt and also seen.  The patterns are made visible through the transmission of light scattered through the membrane at each pixel (Fig.~\ref{fig:concept}F).  The mechanism of actuation, optical powering and addressing technique, simplicity of fabrication, and use of widely available materials render this technology practical and scalable (Table S5, S6).  To demonstrate the ease of manufacturing and scalability, we fabricated devices with 1, 16, 357, and 1,511 addressable pixels (Fig.~\ref{fig:concept}G), using the same fabrication methods for each.  Our results indicate that the devices achieve forces, displacements, resolutions, and response rates meeting perceptual requirements for tactile display, at display scales suitable for diverse applications  (Fig. \ref{fig:concept}H, Fig.~S15, and Table S1). 

\subsection*{Thermo-Mechanical Response Under Photostimulation} 

We experimentally characterized the thermal and mechanical response of individual pixels driven by discrete light pulses with duration $t_p = 50$ ms and power $P_L$. We observed the mean temperature $T(t)$ in the photoabsorber to rise monotonically during exposure (Fig.~2A-D), driving expansion of air within the pixel cavity and deflection of the elastic membrane (Movie~S1). Temperature and displacement both reached their maximum values near the time at which the optical pulse ceased, indicating a nearly static relationship between the absorber and air temperatures during heating (Fig.~2D, Fig.~S8). Following the pulse, temperature and displacement relaxed toward baseline conditions. 

\begin{figure*}[ht!]
    \centering
    \includegraphics[width=163mm]{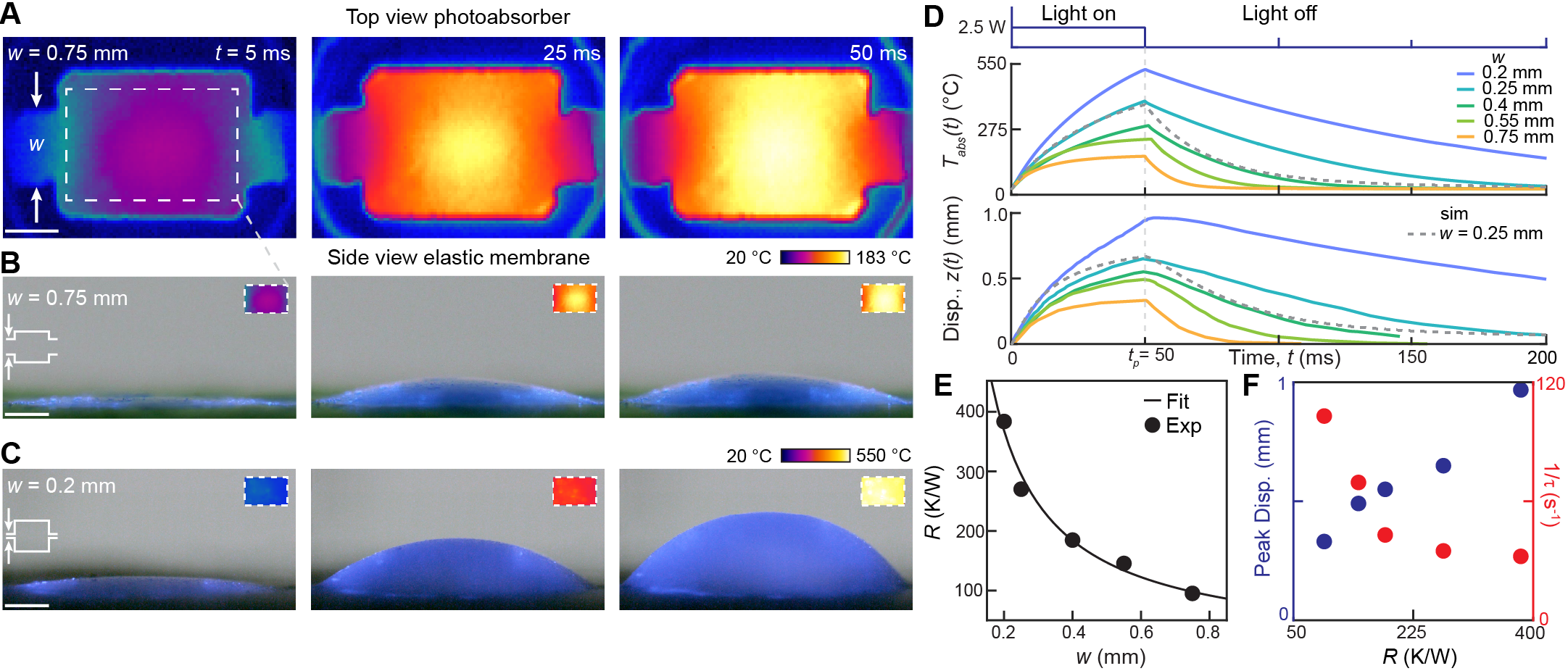}
    \caption{ \textbf{Fig. 2. Thermal-mechanical response under photostimulation.}  (\textbf{A},\textbf{B}) Absorber surface temperature and membrane deflection increased monotonically during illumination with power $P_L = 2.5$ W over 50 ms. (Scale bar = 0.5 mm). (\textbf{C}) Reducing the thermal bridge width from $w=$ 0.75 mm to 0.2 mm yielded higher temperatures and larger displacements. Scale bar: 0.5 mm. (\textbf{D}) Experimental absorber temperature $T(t)$ and membrane displacement $z(t)$ vs.~time $t$ for $w=$ 0.2, 0.25, 0.4, 0.55, and 0.75 mm.   As $w$ decreased, more heat was retained in the absorber, yielding higher temperatures and larger displacements. For $w=0.2$ mm,  at $t=$ 50 ms we observed $T = 527$~$^{\circ}$C and $d = 0.97$ mm.  Grey dashed line: Numerical results for $w = 0.25$ mm; see Fig.~S1. (\textbf{E}) The thermal resistance $R$ of the bridge increased as width $w$ decreased. The data agreed with the theoretically predicted relationship $R = a / w,$ with $a =$ 73.3 mm K/W, $r^2 = 0.98$. (\textbf{F}) Peak membrane displacement and absorber thermal relaxation time $\tau$ increased with thermal bridge resistance $R$, while the relaxation rate $1/\tau$ decreased, indicating that tuning $w$ involves a tradeoff between displacement and relaxation rate.} 
    \label{fig:imaging}
\end{figure*}

The thermal and mechanical response characteristics of the pixel can be tuned by adjusting the photoabsorber geometry. We studied the effect of varying the width, $w$, of the bridges suspending the photoabsorber in the cavity, from $w=$ 0.2 to 0.75 mm (Fig.~2A-D, Movie S2). Larger absorber temperatures, $T(t),$ and displacements, $z(t)$, are obtained from pixels with smaller widths, $w$, indicating that bridge width has a dominant effect on heat transfer out of the pixel. We confirmed this using theoretical and numerical analyses of heat transfer (Text~S1-2). The highest absorber temperatures and displacements were obtained at the smallest width, $w=0.2$ mm. Temperature increased by $507$~$^{\circ}$C and displacement reached a maximum of $0.97$ mm after 50 ms of photostimulation at 2.5 W (Movie S3).  

A thermal resistance parameter, $R$, encodes the influence of photoabsorber bridge width, $w,$ on heat transfer, and can be adjusted to tune pixel response time or displacement. We identified $R$ 
by analyzing the experimental data in tandem with an effective description of heat transfer (Text S1). The time scale $\tau$ governing the absorber temperature $T(t)$ response to photostimulation is given by $\tau = RC,$ where $C$ is the absorber heat capacity.  Values for $R$ and $C$ were obtained by numerical fit to experimental data  (Table~S3).  $R$ is inversely proportional to $w$, in agreement with the theoretically expected relationship $R=L(k_{xy} h w)^{-1}$ where $k_{xy}$ is the in-plane  thermal conductivity of the PGS absorber, $h$ is thickness, and $L$ is the characteristic length of heat transfer (Fig.~2E, Text S1). By adjusting $R,$ we can tune the thermal response time in the range 10 ms $\leq \tau \leq$ 100 ms.  Selecting larger values of $R$ yields larger peak displacements, and longer response times, $\tau$ (Fig.~2F). These competing effects illustrate a trade off between response amplitude and speed that we examined in further experiments.


\subsection*{Rapid Pixel Refresh in Cyclic Photostimulation}

Achieving rapid pixel responses is essential for meeting the perceptual requirements for cyclically refreshed tactile display. Refresh intervals shorter than 50 ms are required in order to ensure perceptual continuity of tactile patterns across actuation cycles \cite{craig1987vibrotactile,sherrick1991vibrotactile}, similar to frame rate requirements for persistence of vision in visual displays. Rapid response rates also enhance the bandwidth of information conveyance and increase display responsiveness.  Our display exhibits rapid actuation cycles, from 2 to 100 ms, that contrast with slower response times, 1 or more seconds, that are often obtained with thermally mediated actuation (Fig.~S15).  This performance is enabled by the high thermal diffusivity of the captive gas, and low thermal mass of the photoabsorber ($100$ $\mu$J/K) and gas ($10$ $\mu$J/K), by the small quantity of heat energy delivered in each pulse (2.5 to 125 mJ), and by favorable heat transfer characteristics, including the surface area to volume ratio, $A/V \propto 1/L$, where $L= 3$ mm is the characteristic pixel dimension. 

\begin{figure*}
    \centering
    \includegraphics[width = 94 mm]{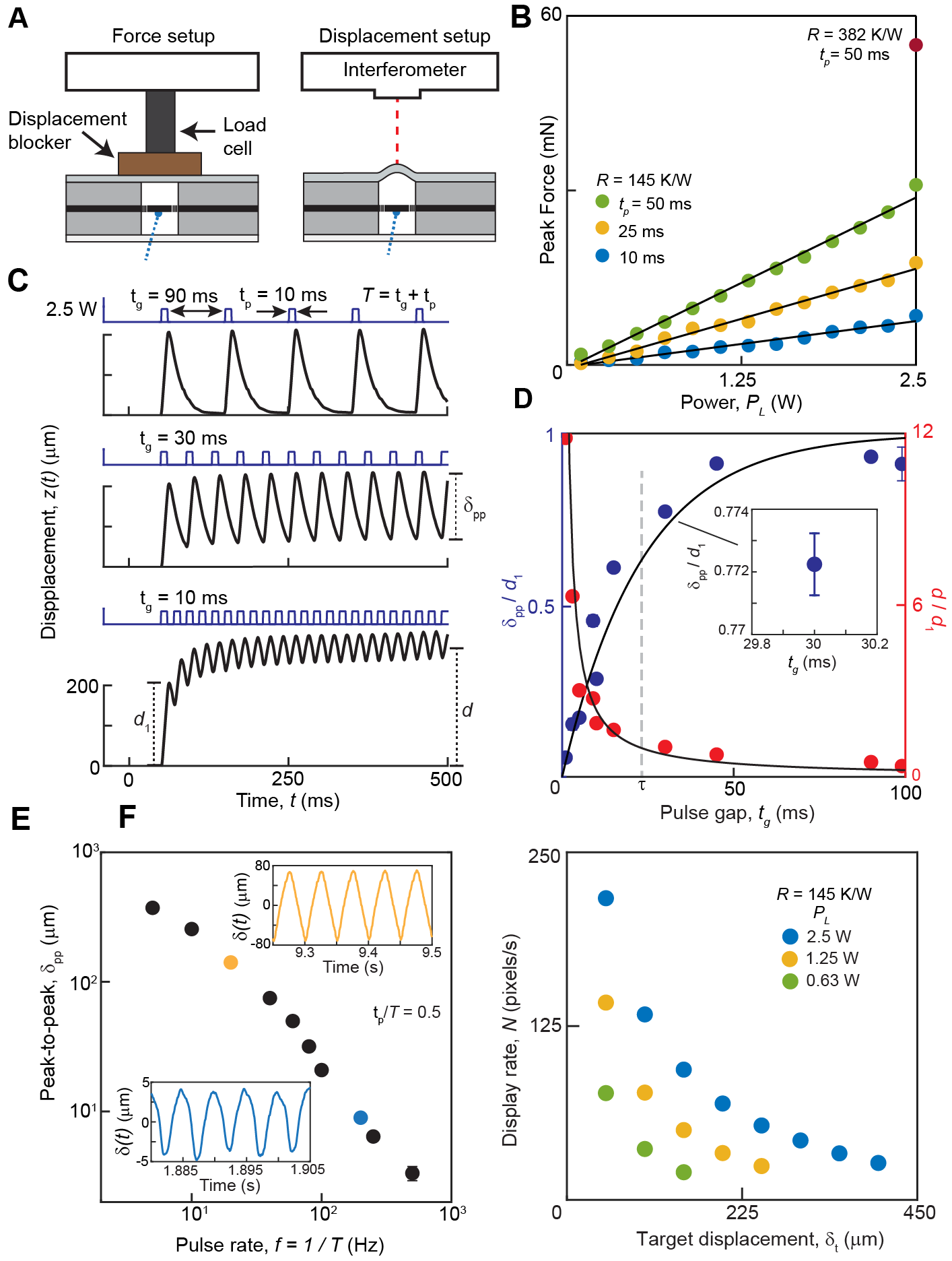}
    \caption{ \textbf{Fig. 3. Cyclic photostimulated actuation.} (\textbf{A}) Isometric force and unloaded displacement measurements.   (\textbf{B})  Individual pixels produced peak forces of $F=$ 55 mN under photostimulation at power $P_L=2.5$ W and pulse duration $t_p=50$ ms (pulse energy 125 mJ).  Force increased linearly with power $P_L$ and increased with pulse duration $t_p$.  (\textbf{C}) Successive displacement responses to pulsed excitation exhibit minimal variation for pulse gap intervals $t_g \gg  \tau$ (Fig.~S7), where $\tau$  is the thermal relaxation time; Here, $\tau \approx 23$ ms and $R = 145$ K/W.  At shorter gap intervals, $t_g,$ heat accumulates, yielding a slowly varying component, $d(t),$ and a reduction in the peak-to-peak oscillation amplitude, $\delta_{pp},$  relative to the first pulse amplitude, $d_1$. (\textbf{D}) As $t_g$ increases, oscillation amplitude, $\delta_{pp},$ (blue) increases; Regression fit:  $\delta_{pp}/d_1 = 1-\exp{(t_g/\tau)}$ ($r^2 = 0.95$). The magnitude, $d$, of the slow component  (red) followed an opposite trend, $d/d_1 = \tau/t_g$ ($r^2 = 0.81$). (Uncertainty bars: 1 SD, $n = 5$ trials). (\textbf{E}) Displacements $\delta_{pp}$ for pulse frequencies $f=$~5 to 500~Hz. Peak-to-peak displacement decreased as $f$ increased, with $t_p/t_g = 1$.  Insets: oscillating component, $\delta(t),$  for $f = 20$ Hz (yellow) and $200$ Hz (blue). (\textbf{F}) In sequential scanning, the maximum pixel display rate, $N,$ is dictated by the required displacement. At fixed $P_L$, $N$ decreases as displacement increases. For $P_L=2.5$ W, $N=$ 217 pixels/s is achieved at 50 $\mu$m, and $N=26.5$ pixels/s is achieved at 400 $\mu$m.}
    \label{fig:mechanical}
\end{figure*}

However, achieving rapid actuation cycles is balanced by the competing need for sufficient dynamic range in force and displacement. This trade-off is parametrized by the effective thermal resistance of the absorber, $R$.  To evaluate this balance in our system, we experimentally characterized the membrane force and displacement responses across a range of optical stimuli (Fig.~\ref{fig:mechanical}).  We observed the peak output force to increase linearly with increasing optical power $P_L$ (Fig.~\ref{fig:mechanical}B).  At constant power, force and displacement increased monotonically with pulse duration $t_p$ (Fig.~\ref{fig:mechanical}B, Fig.~S7).   With  pulse duration held constant, we observed the largest forces ($F=55$ mN for $t_p=50$ ms) at the highest resistance ($R=382$ K/W), and smaller forces ($F=31$ mN) at lower resistance ($R = 145$ K/W; Fig.~3B, S9).

Using these measurements, we computed the stroke efficiency, $\eta_s,$ of actuation as the ratio of stroke power to absorbed optical power, obtaining $\eta=0.03\%$ (Text~S3, Fig.~S11).  
This efficiency is within the range observed for many thermal actuation methods (Table~S1). It reflects several losses. More than 90\% is due to photoabsorber heat transfer to the walls via the bridges (Text~S3).  This thermal conduit can facilitate rapid pixel responses, at the expense of efficiency and dynamic range.  Adjusting $R$ allows us navigate this balance. 
The thermodynamic conversion of heat in the gas to work is about 14\% efficient. The loss represents entropy generation \cite{peterson1999micro}, a measure of irreversibility, and is magnified at small scales. In our device, the pixel volume is $V=7.8$~$\mu$L.  A general scaling analysis indicates that when the characteristic dimension $L$ of a thermally actuated device is sufficiently small, the loss in efficiency scales as $\Delta\eta/\eta_r  \propto L^{-\alpha}$, where $\eta_r$ is the Carnot efficiency, and $\alpha = 1$ or $2$ depending on the heat input configuration (Text~S4). Similar size-efficiency effects have been experimentally observed across many thermodynamic systems \cite{burugupally2018actuatorscaling,sher2009scalinglimitations}.

We next investigated temporal parameters governing the dynamic response of the display under cyclic photostimulation. 
Rapid stimulation produces a multimodal displacement response $z(t) = d(t) + \delta(t),$ with slowly varying and oscillating components, $d(t)$ and $\delta(t)$ respectively 
(Fig.~3C). 
The slow component, $d(t)$, is due to the retention of heat in the cavity across cycles, as occurs when the inter-pulse gap time, $t_g$, is shorter than $\tau.$  The amplitude of $d(t)$ decreases as $t_g$ increases; in steady-state, $d/d_1 = \tau/t_g$  (Fig.~\ref{fig:mechanical}D, red), where $d_1$ is the first pulse amplitude. The amplitude of the oscillating component, $\delta_{pp}$,
exhibits an opposite trend, described by $\delta_{pp}/\delta_1 = 1-\exp(-t_g/\tau)$ (Fig.~\ref{fig:mechanical}D, blue).  Across a range of values of $t_g$, both components fall within the perceptually significant amplitude range \cite{jones2006human}. Thus, by adjusting the pulse gap time, $t_g$, we can control the relative proportion of slow and vibratory tactile feedback. 


Altering the pulse rate, $f=1/(t_p+t_g),$ differentially modulates the amplitude of the two components of displacement.  At constant power and duty cycle, $\nu=t_p/T,$  the amplitude of the slow component, $d$, 
is nearly invariant with respect to variations in pulse rate $f$  (Fig.~S10).  This occurs because the average absorbed power $\overline{P} = \nu \varepsilon P_L$ is determined by the duty cycle and optical power. In contrast, because $t_p$ and $t_g$ both decrease as $f$ increases, the oscillation amplitude, $\delta_{pp},$  decreases (Fig.~\ref{fig:mechanical}E).  Nonetheless, even rapid pulsing at $f=200$ Hz yields an oscillation amplitude of $\delta_{pp} = 8.4$ $\mu$m, which is sufficient to be perceived via touch.\footnote{For comparison, this amplitude is about 25 dB higher than the perceptual detection threshold for 200 Hz vibrations at the fingertip \cite{verillo1971vibrotactile}.} 

While the response rate of cyclically driven pixels is constrained by the thermal response time scale, our display can also render tactile patterns by rapidly scanning the light source across arrays of pixels in sequence.  In sequential scanning, the maximum display rate $N$, in pixels per second, is dictated by the pulse time, $t_p,$ needed to achieve a specified amplitude, $\delta_{pp}$, at the operating power level, $P_L$.  
The maximum pixel rate $N$ thus decreases with increasing displacement (Fig.~3F). At power $P_L=2.5$ W, specifying displacement to be $d=$ 50 $\mu$m results in a maximum display rate of $N=$ 217 pixels/s, while for $d=$ 400 $\mu$m the maximum rate is $N=26.5$ pixels/s.  Higher rates can be achieved by increasing optical power or employing multiple optical sources.

\subsection*{Tactile Pattern Display and Perception} 

\begin{figure*}[ht!]
    \centering
    \includegraphics[width = \linewidth]{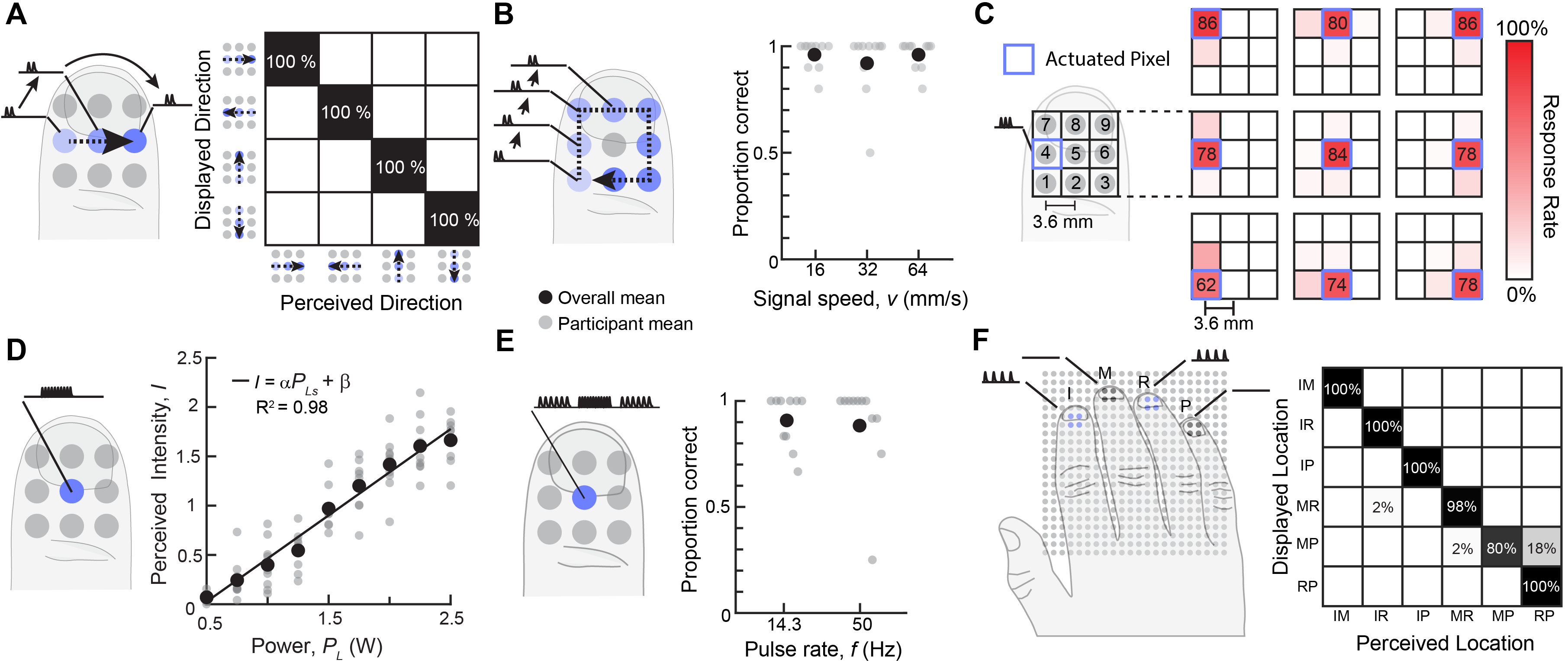}
    \caption{\textbf{Fig. 4. Perception of Dynamic Tactile Pattern Display.} (\textbf{A}) Experiment 1: Perception of linear tactile motion patterns.  All participants correctly reported the motion direction in 100\% of trials.  (\textbf{B}) Experiment 2: Perception of rotational motion for motion speeds $v=$ 16, 32, and 64 mm/s.  Mean response accuracy: 94.7\%. Black dots show the overall mean, and grey dots show participant mean. (\textbf{C}) Experiment 3: Spatial localization of actuated pixels near the finger.  Mean accuracy: 78.4\% (99\% within 1 pixel distance).  Mean localization error: 0.17 mm. (\textbf{D}) Experiment 4: The perceived intensity of pulse train feedback from a single pixel increased linearly with optical power $P_L$ ($\alpha = 0.022$, $\beta = -0.42$, $r^2 = 0.98$).  Stimulus parameters: $N=20$ pulses, $t_p=t_g=$ 25 ms, $0.5 < P_L < 2.5$ W.   (\textbf{E}) Experiment 5: Discrimination of temporal patterns.  "Odd-One-Out" task to identify which of three sensations was different. Total accuracy: 89.6\%. 
    (\textbf{F}) Experiment 6: Perception of multi-point tactile patterns felt at any two of four fingers.  Participants reported the locations that were stimulated with 96.3\% accuracy.}
    \label{fig:perceptual}
\end{figure*}

This tactile display employs a physical operating principle that diverges from those used in prior haptic technologies. The forces and displacements that can be produced by individual pixels (up to 55 mN and 0.97 mm) are well within the perceivable range \cite{jones2006human}. These basic capabilities led us to investigate the potential of using multiple pixels to render a variety of perceptually distinct tactile patterns.  
We conducted six experiments assessing the perception of linear and rotational motion patterns, tactile localization, the scaling of perceived intensity with power, the perception of temporal patterns, and multi-finger tactile perception (Figs.~4A-F).  The same ten participants completed all six experiments without  training.

Experiment 1 assessed the perception of linear motion patterns that were generated through the sequential activation of pixels along each of four cardinal directions (Fig.~\ref{fig:perceptual}A). All participants correctly reported the motion direction in 100\% of trials.  In experiment 2, participants discriminated rotational motion patterns at three different motion speeds (Fig.~\ref{fig:perceptual}B).  Their response accuracy was 94.6\%.  
These results underscore the effectiveness of the display in representing motion.

In experiment 3, participants reported which single pixel was actuated out of nine pixels arranged in a 1 cm $\times$ 1 cm grid beneath the finger pad 
(Fig.~\ref{fig:perceptual}C).  They responded correctly on 78.4\% of trials. The mean error in localization was 0.17 mm, with 99\% of all responses within one pixel of the correct location.  As a point of comparison, the two point discrimination threshold, a standard measure of tactile acuity, averages 2 to 3 mm at the fingertip \cite{cellis1977two}. 

In experiment 4, we assessed the relationship between perceived intensity and optical power (Fig.~\ref{fig:perceptual}D), using the psychophysical method of magnitude estimation. Participants rated the intensity of impulse trains that were excited at different power levels (0.5 to 2.5 W). Perceived intensity increased linearly with optical power ($r^2 = 0.98$), and thus increased linearly with force (Fig.~\ref{fig:mechanical}B).  Thus, perceptually distinct levels of tactile feedback can be generated by modulating power.   

In experiment 5, participants discriminated temporal patterns  (Fig.~\ref{fig:perceptual}E), using an odd-one-out paradigm.  In each trial, they felt a sequence of three pulse train stimuli generated at a single pixel.  Two stimuli had the same pulse rate, either slow or fast, 
while the third stimulus -- the oddball -- had a different pulse rate (respectively fast or slow). 
Participants correctly discriminated the temporal patterns, identifying the oddball stimulus, in 89.6\% of all trials. 

In experiment 6, we investigated the perception of multi-point tactile pattern perception. In each trial, participants simultaneously felt stimuli on two out of four fingers in contact with the display, and reported the pair of stimulated locations. Participants correctly identified the pair of locations with 96.3\% accuracy. 

In post-hoc surveys, no participants reported perceiving thermal sensations or seeing any optical stimuli. 

\section*{Conclusion}

Our tactile display technique exploits the advantages of projected light for power transmission, addressing, and actuation.  Light is directly transduced into forces and displacements via energetically passive surfaces populated with optotactile pixels.  The actuation mechanism, layered architecture, and simplicity of fabrication make it straightforward to produce displays with dozens to thousands of pixels, with similar investments in material cost and fabrication time.  We present implementations with up to 1,511 pixels -- significantly more than any previous device capable of similar performance, notably in response time and amplitude.  Our technique furthermore yields tactile patterns that can be not only felt, but also seen, enabling multisensory tactile displays.  These displays hold promise for applications ranging from automotive interfaces to virtual reality controllers. In order for them to achieve widespread adoption, higher pixel display rates are needed.  This could be achieved using optimized architectures that exploit fundamental knowledge about thermodynamic device efficiency, and by leveraging high-power optical projection technologies like those being developed for additive manufacturing and automotive applications \cite{somers2024physics}. These improvements could enable widespread deployment of high-definition projected light tactile displays that lend physical form to otherwise intangible digital data.

\section*{Materials and methods}

\subsection*{Display fabrication}
The same manufacturing steps were used for displays with 1 to 1,511 pixels. The display surface is a layered assembly. In order from base to top layer, the materials are: Transparent acrylic sheet, 
one-side adhesive polysiloxane (PS) sheet (thickness 0.5 mm), 
thick pyrolytic graphite sheet (thickness 17 $\mu$m), 
one-side adhesive PS sheet, cast silicone elastomer (thickness 0.25 mm).  
Double-sided adhesive tape (2477, 3M, USA) was used to affix the lower PS sheet to the acrylic. The thickness of the acrylic may be selected based on application requirements. For the flexible display, the thickness was 0.01", and for the rigid display, it was 1/16". Supplier information, and representative costs for the materials, are reported in Table S5.

The PS sheets were patterned via laser cutting to form a grid of cylindrical cavities  ($r_\mathrm{cavity}$ = 1.5 mm, pitch = 4 mm; Fig. S5). The PGS was patterned using a vinyl cutter to form a grid of rectangular photoabsorbers (2 mm x 1.5 mm) aligned with the cavities (Fig. S5). 
The acrylic was laser cut to form the square geometry; for the 1,511 display, the dimensions are 162 mm x 162 mm (Fig. S5). The PGS adhesive was dissolved via acetone processing, and aligned between PS layers.  For the cast elastomer, the two silicone elastomer component reagents were mixed, degassed, and placed into a mold with depth 250 $\mu$m . It was subsequently cured at room temperature at room temperature for 24 hours. Because the resulting elastomer sheet is tacky, no adhesive was needed to affix it to the PS layer.

\subsection*{Photoabsorber Thickness and  Density Characterization}

The thickness of the PGS photoabsorber was measured via profilometry and confocal microscopy.  For the profilometer measurement (DektakXT, Bruker), the adhesive side of the PGS was applied to a glass slide.  Measurements from four distinct samples of the combined thickness of PGS and adhesive were taken, yielding  $h_{PGS,adh} = 23.2 \pm 1.2 \mu$m (Fig. S16). We characterized the adhesive thickness using confocal microscopy (DSX1000, Olympus), obtaining $h_{adh} = 6.5 \pm 1 \mu$m.  Subtracting the two thicknesses, yielded the PGS photoabsorber thickness, $h_{PGS} = 16.7 \pm 2.2 \mu$m. The result was 6.7 $\mu$m larger than the thickness of 10 $\mu$m specified by the manufacturer. The PGS surface roughness was $\pm 1.5$ $\mu$m.  
The mass density of the PGS material, after adhesive removal, was determined using a microbalance (XPR105, Mettler Toledo), and dimensional measurements of surface area ($25 \pm 5$ mm$^2$) and thickness $h_{PGS}$ (as noted above).  This yielded a PGS density value of $\rho_{PGS} = 1962 \pm 650$  kg/m$^3$, which is consistent with the manufacturer specified value of 2130 kg/m$^3$.

\subsection*{Optical stage}
The optical stage included a 2-axis galvo mirror system (Laserwinkel, Netherlands), a 4W 450 nm diode laser (LT-20W-A, Laser Tree), a collimating lens (Qiaoba), and an F-theta lens (Opex). The optical components were arranged in a 3D printed fixture. The laser path proceeded through the collimating lens, galvo, F-theta lens, and through the optical window of the addressed pixel, onto the respective photoabsorber  (Fig. S6). The optical stage was configured in an optically sealed enclosure with an aperture located at the top, where the display surface is positioned. Representative costs of  components used to construct  the optical stage are given in Table S6.

The laser was powered with 12 V using a laboratory supply.  Optical power output was modulated by PWM signal (PWM frequency: 5 kHz) supplied to the laser by  a data acquisition unit (NI SCB-68A) under computer control. 

The maximum instantaneous optical power incident on the sample was 2.45 W, as measured  via bolometer (S212A, Thorlabs, USA).  This measured power accounted for transmission losses through the mirrors and lenses. Transmission through the acrylic layer was 92.4\% at 450 nm (Fig.~S2), as determined by spectroscopic measurement (UV-1800 UV-VIS, Shimadzu, Japan). The optical absorption of the PGS at 450 nm was 72.3\% (Fig.~S2).  At maximum optical power, the power absorbed by the PGS was 1.63 W, accounting for all of the aforementioned losses.

\subsection*{Imaging}

The PGS photoabsorber material and geometry were characterized using a microscope (AmScope, USA) and a laboratory scale with $50$ $\mu$m divisions. 
We used a MWIR camera (Fast M3K, Telops, USA) and microscope lens (1x, Telops, USA) to collect the thermal data in Fig. 2. The camera was directed at the sample via a mid-IR gold mirror (PFSQ20-03-M02, Thorlabs, USA) (Fig. S14A, B). The top elastomer sheet of the display was removed during thermal imaging. Time-resolved infrared videos were captured at a frame rate of 5,000 fps using an exposure time of 5.04 $\mu$s, and subsequently analyzed in RevealIR (Telops, USA) where an emissivity of 0.39 was used (Fig. S3). 
Radiometric temperature calibration was performed separately for low and high-temperature ranges, $T<236 ^{\circ}$C and $T>236 ^{\circ}$C.  Data captured across both ranges was combined in software (Fig. 2D), and aligned via temporal synchronization. For the sample with $w = 0.55$~mm, we performed model-based calibration (Eq. (8) in Text S1) to infer temperatures $T> 236 ^{\circ}$C.

Video data was captured using a high-speed camera (Phantom VEO 640L, Vision Research, USA) with a microscope objective and a 2x Plan Apo Infinity Corrected lens (WE1603, Mitutoyo, USA). The camera was directed at the profile of each sample (Fig.~S14C, D). A measured sample (thickness 2.2 mm) of cast acrylic was introduced into the field of view as a dimensional scaling reference. Optical imaging data were collected, reviewed, and saved in Phantom Video Player (Vision Research, USA).  Metrology analysis was performed via manual annotation of points of interest using video analysis software (Tracker, Version 6.1.5), yielding displacement values (Fig. 2D-F).

An extended exposure photo, used for illustration in Figure 1, was captured using the camera of an iPhone SE 3rd generation via the app Spectre, with the "light trail" feature, turned on. This image was not used for quantitative analysis.

\subsection*{Thermal resistance measurements}
Values representing net effective thermal resistance $R$ to heat transfer in and out of the absorber, and absorber heat capacity $C$, were obtained from experimental data, by regression fit to the solution of a heat transfer equation (Text S2), with timescale $\tau = R C$.  The fit was performed using temperature data acquired during the first 30 ms of photostimulation at 1.63 W. 
The temporal dependence of temperature, through $\tau = RC$, and the value of $R$ obtained from the fit, is generic and insensitive to modeling choices.  The absorber was initially at room temperature during data collection.  The experimentally determined heat capacity  was $C = 101 \pm 16$ $\mu$ JK$^{-1}$, which was in agreement with the value we calculated based on the geometry and properties of the photoabsorber, using $C = V_{abs}\rho_{abs}c_{abs} = 85$ $\mu$ JK$^{-1}$, where $V_{abs}$, $\rho_{abs}$, and $c_{abs}$ are the volume, density, and specific heat capacity of PGS optical target, respectively.

\subsection*{Numerical experiments}
We studied the thermal and mechanical response characteristics of our system using finite element numerical analysis (FEA, COMSOL Multiphysics). The simulations accounted for heat transfer, gas expansion, and membrane mechanics. Detailed descriptions of the experiments and simulation methods are documented in Text S2.

\subsection*{Mechanical characterizations}
Force measurements were captured with a load cell (LSB200 FSH02602, Futek, USA) and a strain gauge amplifier (CSG110, Futek, USA), using a data acquisition unit.  The sample rate was 10 kHz. The load cell was applied with a preload of 0.74 N to ensure stable measurement conditions. Surface displacement measurements, as reported in Fig. 3, were captured via Laser Interferometer (Mikrotrak II LTC-025-02, MTI Instruments, USA) at a sample rate of 100 kHz. 

\subsection*{Perceptual experiments}

All six perceptual experiments were performed using a 437-pixel display surface (7.5 cm x 7.5 cm), with a pixel pitch of 3.6 mm (pixel density: 7.7 cm$^{-2}$).

To inhibit light leakage, the flexible membrane (EcoFlex 00-10) was dyed white (Silc Pig White, Smooth-On), and a sheet of gold leaf ($\sim$0.1 $\mu$m thick) was adhered to the base.  Neither modification significantly affected the membrane mechanical response. A black curtain occluded the display from participants' view during the experiment. While we observed no stray light, participants wore laser safety glasses to avoid safety concerns.

All participants gave their written informed consent.  The protocol was approved by the human subjects review board at the authors' institution.  The same ten participants (4 female, 6 male, age 19 to 34 years) completed the sequen ce of six experiments without training. None reported diagnosed health issues or ailments affecting touch sensation.  One additional participant was unable to complete the perceptual tasks for personal reasons unrelated to the experiment or device.  This data was excluded.  Participants' hand dimensions were measured. Right index finger pad widths were between 1.4 to 2.0 cm (mean: 1.6 cm), and hand lengths were between 16.5 to 19.5 cm (mean: 18.3 cm). Participants wore noise isolating circumaural headphones playing white noise to mask auditory cues during the experiment. 
For each experiment, participants finger were positioned as depicted in Figure 4. 
Before each experiment, participants were presented with each stimulus once for familiarization.   Detailed descriptions of the design, method, protocol, and analyses used for each of the six perceptual experiments are reported in Text S5.


\clearpage
\newpage

\small
\bibliographystyle{Science}

\newpage
\normalsize
\section*{Acknowledgments}
We gratefully acknowledge Yangying Zhu, Shantal Adajian, Bolin Liao, and Alban Sauret for assistance with characterization experiments. We thank Neeli Tummala for valuable comments. We acknowledge the Innovation Workshop and Microfluidics Lab in the UCSB California NanoSystems Institute, the UCSB MRSEC facility (NSF DMR 0520415), and the UCSB Nanofabrication facility, which were used in this research. 
\newline

\noindent \textbf{Author Contributions:} M.L. and Y.V. designed research; M.L, G.R., D.G. and Y.V. performed research; M.L. and V.K. analyzed data; M.L., E.H., and Y.V. wrote the paper.
\newline

\noindent \textbf{Competing Interests:} The authors declare that they have no competing interests.
\newline

\noindent \textbf{Data and Materials Availability:} All data needed to evaluate the conclusions in the paper are present in the paper and the Supplementary Materials. These data are also available on Zenodo (10.5281/zenodo.13948334).
\newline

\noindent \textbf{Supplementary Materials:}\\
Supplementary Text\\
Figs. S1 to S17\\
Tables S1 to S6\\
References (35 - 76)\\
Movies S1, S2 and S3\\

\clearpage

\section*{Supplementary texts}

\subsection*{S1. Thermodynamic modeling}

We used an effective mathematical description of heat transfer in the optotactile pixels to gain insight into factors influencing their thermal response.  The model accounts for heat transfer from the absorber to air and to the surrounding wall region.  The solutions enabled us to identify the effective thermal parameters $R$ and $C$ representing net effective thermal resistance for heat transfer through the absorber and absorber heat capacity.  We used the ideal gas law to relate temperature to gas pressure.  A linear elastic model was sufficient to capture the membrane displacement, as confirmed by comparison with experimental and numerical data. 

\subsubsection*{Heat transfer analysis}    

\begin{figure*}[b!]
    \centering
    \includegraphics[width = \linewidth]{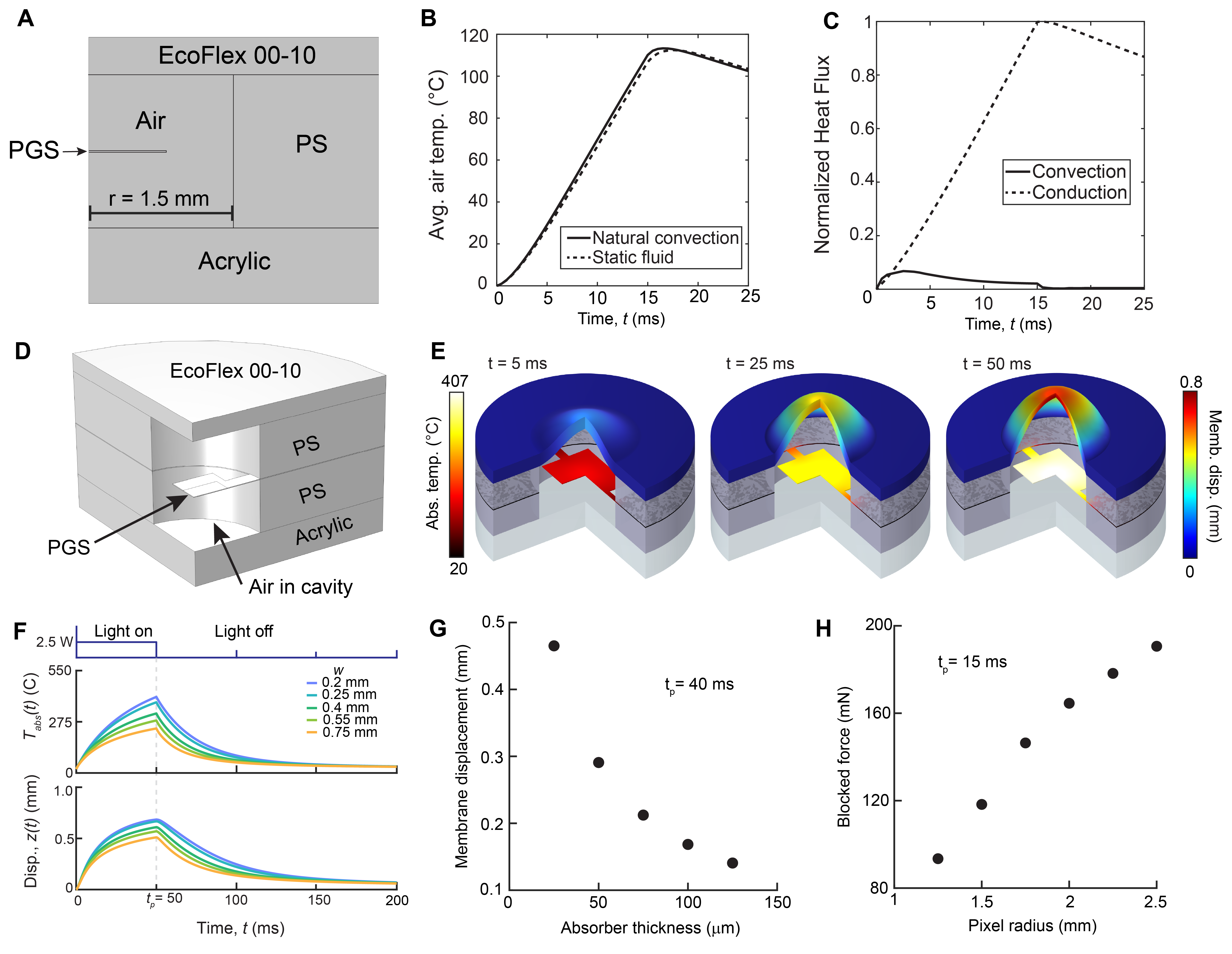}
    \caption{\textbf{Fig. S1. Numerical Experiments} A) The geometric and material setup for the 2D FEA simulation. B) Nearly identical mean cavity air temperatures were obtained using an effective heat transport model that accounted only for air conduction and using a full fluid dynamics simulation accounting for convection. C) Heat transport in the gas was primarily governed by conduction, with convection having a notable influence only within the first 1 to 2 milliseconds of heat application. D) Geometric and material setup for 3D pixel simulation. E) Snapshots of the FEA model at t = 5, 25 and 50 ms for $w = 0.4$ mm. The left color bar corresponds to the absorber temperature, whereas the right indicates membrane displacement. F) Time-resolved temperature and displacement curves for $w = $ 0.2, 0.25, 0.4, 0.55, and 0.75 mm. G) The membrane displacement decreases as absorber thickness, $h$, increases. H) The blocked force increases as the pixel radius increases.
    }
    \label{supp:fea}
\end{figure*}

We developed an effective description of heat transfer with state variables given by the mean (spatially averaged) absorber temperature, $T_{abs}(t)$, and air temperature, $T_{air}(t)$.  An equivalent circuit illustration is given in Fig.~S\ref{supp:analytic model}A.  Heat transfer is described by the following coupled differential equations.
\begin{equation}
    C_{abs}\frac{dT_{abs}(t)}{dt} = \varepsilon P_L - \frac{T_{abs}(t)-T_{air}(t)}{R_{air}}-\frac{T_{abs}(t)-T_{wall}}{R_{abs}}
    \label{eq:abs_diff_eq}
\end{equation}
\begin{equation}
    C_{air}\frac{dT_{air}(t)}{dt} = \frac{T_{abs}(t)-T_{air}(t)}{R_{air}}-\frac{T_{air}(t)-T_{wall}}{R_{air}} 
    \label{eq:air_diff_eq}
\end{equation}
Here, $C_x$ is the thermal mass for material with label $x$, $P_L$ is the incident optical power, $\varepsilon$ is the absorption coefficient, and $T_{wall}$ is ambient temperature. $R_{air}$ is the thermal resistance along the $z$ direction, to a distance $H$ equal to the cavity height. $R_{abs}$ is the thermal resistance for heat transfer from the absorber to the surrounding wall region, via the absorber bridge features of width $w$. Due to the small cavity dimensions, heat transfer from natural convection may be neglected (SM text S2, Fig. S\ref{supp:fea}B and C). While the solutions to the coupled equations can be obtained analytically, a scaling analysis shows that they approximately separate because heat transfer out of the absorber occurs predominantly through the bridge features.  Thermal resistance through the bridge features is of order

\begin{equation}
    R_{abs} \sim \frac{L}{k_{abs}wh},
\end{equation}
while thermal resistance through the air is of order
\begin{equation}
    R_{air} \sim \frac{H}{k_{air}A_s} 
\end{equation}
Here, $w, L,$ and $h$ are the absorber bridge length, width, and thickness, respectively, $k_{abs}$ is the thermal conductivity of the PGS material, and $A_s$ is the photoabsorber surface area. Comparing these for our system quantities yields the ratio
\begin{equation}
    \frac{R_{abs}}{R_{air}} \sim \frac{L}{H}\frac{k_{air}}{k_{abs}}\frac{A_s}{wh}  \sim 10^{-4}
\end{equation}

Thus, heat transfer out of the absorber is dominated by the pathway through the absorber bridge features. Consequently, the term in equation (\ref{eq:abs_diff_eq}) representing heat transfer through the air can be neglected, separating the differential equations

\begin{equation}
    C_{abs}\frac{dT_{abs}(t)}{dt} = \varepsilon P_L-\frac{T_{abs}(t)-T_{wall}}{R_{abs}}
    \label{eq:abs_diff_eq_uncoupled}
\end{equation}
\begin{equation}
    C_{air}\frac{dT_{air}(t)}{dt} = \frac{T_{abs}(t)-T_{air}(t)}{R_{air}}-\frac{T_{air}(t)-T_{wall}}{R_{air}} 
\end{equation}
The solution to equation (\ref{eq:abs_diff_eq_uncoupled}) from the initial condition $T_0$ is
\begin{equation}
    T_{abs}(t) = \varepsilon P_LR_{abs}\left(1-e^{-t/\tau_{abs}}\right) + T_0e^{-t/\tau_{abs}}
    \label{eq:absorber-temp}
\end{equation}
where $\tau_{abs} = R_{abs}C_{abs}$ is the characteristic time constant over which heating occurs.  This analysis shows that the net effective thermal resistance for heat transfer from the optical source through the absorber is approximately $R_{abs}.$  Thus, in the remaining analysis, and in the main text, we use the variables  $R=R_{abs}$ and $C=C_{abs}$ to represent the effective thermal resistance and heat capacity, and $\tau=\tau_{abs}$ to represent the corresponding time scale.

\subsubsection*{Air temperature}

We consider $T_{abs}(t)$ to be an arbitrary (specified) input signal.  The air temperature satisfies
\begin{equation}
    C_{air}\frac{dT_{air}(t)}{dt} = \frac{T_{abs}(t)-T_{air}(t)}{R_{air}}-\frac{T_{air}(t)-T_{wall}}{R_{air}} 
\end{equation}
Let the wall and air be initially at ambient temperature, $T_{wall} = T_{air}(t=0) = T_0$, and let $\tau_{air} = R_{air} C_{air}$. The solution is
\begin{equation}
    T_{air}(t) = \frac{1}{\tau_{air}} \int_0^t T_{abs}(\tau) e^{-2(t - \tau)/\tau_{air}} d\tau \ + \ \frac{T_0}{2} \left(1 + e^{-2t/\tau_{air}}\right)
    \label{eq:soln4}
\end{equation}
Thus, changes in absorber temperature $T_{abs}(t)$ drive changes in air temperature, yielding gas expansion over a timescale of $\tau_{air}/2$, where $\tau_{air} = R_{air} C_{air}$. 

\subsubsection*{Pressure, Force, and Displacement}
In isometric conditions, pixel cavity volume is constant, the pressure, $P(t)$, is given by 
\begin{equation}
    P(t) = \rho R_s T_{air}(t),
\end{equation}
where $\rho$ is the gas density (air density in our experiments), and $R_s$ is the specific gas constant.  Spatial variations in cavity pressure can be neglected, because the thermal time constants governing heat transfer are much longer than the characteristic acoustic time scale in the cavity: $\tau_{abs}, \tau_{air} \gg L / v \approx 10^{-5}$ seconds, where $v$ is the speed of sound and $L = 3$ mm is the characteristic cavity dimension.

The net force on the elastic membrane is $F(t) = A_p \cdot P(t),$ where $A_p = \pi r^2$ is the pixel aperture surface area at the membrane.  We observed the membrane deflection $z(t)$ to be adequately captured by a linear elastic model (Fig. \ref{supp:ecoflex-linearity}).  
\begin{equation}
    z(t) = \frac{3}{1280} \frac{(1-\nu^2)P(t)r^4}{Eh_m^3}
\end{equation}
Here, $\nu$ is the membrane material poisson ratio, $E$ is Young's modulus, and $h_m$ is the membrane thickness.   Substituting, 
\begin{equation}
    z(t) = \frac{3}{1280} \frac{(1-\nu^2)\rho R_s T_{air}(t) r^4}{Eh_m^3}
\end{equation}
The time-dependent air temperature, $T_{air}(t)$ as driven by  absorber temperature $T_{abs}(t)$ is governed by equation (\ref{eq:soln4}). Thus, the membrane displacement is proportional to $T_{air}(t)$, and responds to changes in $T_{abs}(t)$ with linear dynamics governed by (11). 









\subsection*{S2. Finite element simulations}

\subsubsection*{Simulation Design}
We studied the thermal and mechanical response characteristics of our system using finite element numerical analysis (FEA, COMSOL Multiphysics). The simulations accounted for heat transfer, gas expansion, and membrane mechanics.  Material parameters and geometry matched our device (Table S4 and Fig.~S1). 
PGS is known for its high in-plane thermal conductivity $k_{xy}$, reported by the manufacturer to be  1950 W/m$\cdot$K \cite{Panasonic_PGS_2024}. We independently characterized thermal conductivity of our PGS samples at $T=300$ K using a previously developed steady-state vacuum measurement technique  \cite{zhao2016conductivitymeasurement}. We extrapolated from this value, using the $1/T$ dependence of PGS thermal conductivity on temperature \cite{null1973graphite}, obtaining $k_{xy}(T) = 3.6\times 10^5 / T$ W/(m$\cdot$K); near $T=300$ K, the thermal conductivity is $k_{xy} = 1.2\times 10^3$ W/(m$\cdot$K).  Cross-plane thermal conductivity was set to $k_z = 15$  W/(m$\cdot$K) based on manufacturer specifications, which we independently experimentally confirmed.  The thermal conductivity in the wall region was $k_{xy,wall} = \alpha \, k_{xy}$, where the value $\alpha = 0.091$ was experimentally determined to account for unobserved variables affecting heat transfer (Text S2). The membrane material (EcoFlex 00-10) was treated as linear elastic, which is justified over the range of stresses and strains in this work \cite{marechal2021} (Fig.~S13).  

Heat, representing absorbed optical energy, was supplied using a boundary heat source spanning a rectangular region on the bottom face of the PGS whose dimensions (1 mm $\times$ 0.5 mm) matched the rectangular focus of our diode laser.  Input power was matched to our experimental conditions.  Additional numerical studies confirmed that the fluid could be treated as static (Text S2, Fig.~S1B), and that thermal convection could be neglected (Fig.~S1C).  Gas (air) pressure was determined by the ideal gas law,  
\begin{equation}
    P(t) = P_{atm} \left[ \frac{\overline{T}(t)}{T_{atm}} \frac{V_{0}}{V(t)} - 1\right]
\end{equation}
Here, $\overline{T}(t)$ is spatially averaged air temperature, $T_{atm}=300$ K was ambient temperature, $P_{atm}$ was ambient pressure, $V_0$ was initial cavity volume, and $V(t)$ was cavity volume accounting for membrane deflection. 

\subsubsection*{Simulation Results}
In two and three-dimensional simulations, with and without natural convection (Fig.~S1A), we observed similar responses upon deposition of heat in the photoabsorber for $t_p = 15$ ms.  Consistent with the foregoing analysis, we obtained nearly identical mean cavity air temperatures upon simulation with a simpler model of heat transport accounting for conduction and convection as we did when performing a full fluid dynamics simulation (Fig.~S1B). Heat transport in the gas was dominated by conduction, with convection playing a significant role only during the first 1 to 2 ms of heat application (Fig.~S1C). This aligns with theoretical predictions since the cavity features are of the same length scale as the momentum boundary layer of the fluid \cite{bennett2012transportphysics}.  The gas was thus assumed to be still for the remainder of the numerical studies. 

Simulation parameters matched our device geometry and manufacturer supplied or experimentally determined parameters (see Methods).  In-plane heat transfer through the laminated assembly of the walls bounding the pixel cavity depends on unobserved variables including contact conditions and thermal properties and temperature of the adhesive and PDMS.  We accounted for these effects by matching simulation output to our experimental data, which yielded  an effective value of thermal conductivity in the wall region, $k_{xy,wall}(T) = \alpha \, k_{xy}(T),$ with $\alpha = 0.091.$  

Numerical results obtained from 3D FEA simulation of pixel responses (Fig.~S1D) were in agreement with our experimental observations. The addition of heat led to rapid temperature increase in the optical absorber, and drove rapid gas expansion and membrane deflection (Fig.~S1E). We studied the effect of bridge width, for values $w = 0.2$ mm to $0.75$ mm.  The range of peak average absorber temperatures we observed  (237 to 402 C) was somewhat compressed relative to our experimental measurements (162 to 527 C), while numerically determined displacements (0.51 to 0.68 mm) were likewise somewhat compressed in magnitude (0.33 to 0.97 mm) (Fig.~S1F). These differences may be attributable to unmodeled characteristics, such as variations in absorber microgeometry, membrane dynamics, or unobserved variables affecting heat transfer  \cite{butland1973graphite,null1973graphite}. 

We performed numerical experiments with parametric variation of absorber thickness, observing decreasing membrane displacement with increasing absorber thickness (Fig.~S1G). This finding may be attributable to two effects. First, increasing absorber thickness reduces the thermal resistance, $R$, of the bridges connecting the optical absorber to the walls, causing more of the heat to bypass the gas. Second, increasing absorber thickness increases the absorber thermal mass, $C$, and thus reduces the temperature increase of the absorber produced by the addition of heat $\Delta Q$. Informed by this finding, and by the high opacity of PGS, we selected the thinnest commercially available PGS material for our experimental devices.  

We studied the effect of increasing the radial scale of the pixel cavity, observing that as the radial dimensions of the device increase, force output increases  (Fig.~S1H), as can also be inferred from simple theoretical considerations. However, the pixel cavity radius constrains the pixel density, and thus the display resolution.  Selecting the pixel cavity radius to be  $r = 1.5$ mm provided sufficient pixel density (approx.~3 mm linear pitch,  $\sim$8 pixels/cm$^2$), and enabled the production of a variety of forces within the range that can be perceived (Fig.~4D).

\subsection*{S3. Energy and Efficiency}

Energy is supplied to our display device via targeted light transmission.  Our characterizations indicated that the proportion of light energy converted to heat in the photoabsorber is $\varepsilon = 66\%$.  These optical energy losses, 34\%, which are due to optical absorption in the acrylic window and photoabsorber reflectance, could be reduced through the use of an optical window with higher transmittance, and through reductions in reflectance that could be realized by absorber material processing or selection.   

To compute the stroke efficiency, $\eta_s$, of the pixels, we used our experimental data to calculate a measure of mechanical energy output --  stroke work -- which is given by $W_s = \frac{Fd}{2}$,  where $F$ is the peak force of the membrane and $d$ is the peak displacement.    This measure of mechanical energy output is somewhat idealized, but has been used in prior studies \cite{heisser2021combustion,aubin2023combustion}, facilitating comparison.  We calculated output mechanical power as stroke power, $P_s = \frac{W}{\Delta t}$, where $\Delta t = t_p$ was the duration of photoexposure.\footnote{$t_p$ approximated the duration over which the membrane deflection increased, because $\tau_{air}\l \tau_{abs}$.}  Stroke efficiency was obtained as the  ratio $\eta_s = \frac{P_s}{\varepsilon P_L}$ of $P_s$ to absorbed optical power, $P_L$.  The maximum efficiency  $\eta_s = 0.03 \%$ was obtained using a pixel with $R = 382$ J/K, with $t_p=50$ ms  (Fig.~S11). The value of $\eta_s$ we obtained is in the range of efficiencies observed for many thermally actuated devices (Table~S2). 
We next estimated the efficiency of heat energy transfer from absorber to gas, $\eta_{qt},$ and from heat in the gas to mechanical work output, $\eta_{tm}$.  To compute these, we used the measured force values, $F(t),$ to compute cavity pressure $P(t)$, and applied the ideal gas law to compute the change in air temperature, obtaining $\Delta T_{air}(t) = 20.5$ C. This implied a change in the internal energy of air of $\Delta Q_{air} = 0.18$ mJ.   Dividing this quantity by energy absorbed yielded an efficiency of heat transfer from absorber to gas of $\eta_{qt}=0.22\%$.  This value implies that more than 90\% of absorbed energy is lost to photoabsorber heat transfer to the walls via the bridge features.  The thermal conduit of the bridges speeds the pixel response time, but decreases efficiency.  In our design, adjusting $R$ allows us to tune performance, trading off efficiency and speed.  These losses could be greatly reduced via optimized architectures that exploit well-established thermodynamics knowledge and optimized designs for thermomechanical energy conversion efficiency in analogous devices \cite{burugupally2018actuatorscaling}. This analysis also implied that the efficiency of thermomechanical energy conversion by the gas is $\eta_{tm} = 14.3\%$.  The corresponding loss is primarily due to entropy generation \cite{peterson1999micro}, a measure of irreversibility, which is magnified at small scales. In our device, the pixel volume is $V=7.8$~$\mu$L.  A theoretical scaling analysis provided in Section S4, below, indicates that when the characteristic dimension $L$ of a thermally actuated device is sufficiently small, the relative loss in efficiency $\eta$ scales as $\Delta\eta/\eta_r  \propto L^{-\alpha}$, where $\eta_r$ is the Carnot efficiency, and $\alpha = 1$ or $2$ reflects the thermodynamic design  (Text~S4). Similar size-efficiency effects have been experimentally observed across many thermodynamic systems \cite{burugupally2018actuatorscaling,sher2009scalinglimitations}.

\subsection*{S4. Dimensional Scaling Effects on Thermodynamic Efficiency}

The pixels operate based on a thermodynamic principle, converting heat from incident light to mechanical work.  The magnitude of the experimentally determined efficiency of our device, $\eta = 0.03\%,$  is consistent with efficiencies reported for many thermally mediated actuators.  It also reflects adverse effects of dimensional scaling on the efficiency of thermodynamic microactuators, due to the small volume ($V=7.8$ $\mu$L) of our pixels.  

To illustrate the effect of scaling on efficiency in devices such as ours, consider a model thermodynamic actuator with characteristic dimension, $L$.  Efficiency may be written as $\eta = 1 - Q_{out}/Q_{in} = \eta_r - \Delta \eta$, where $\eta_r = 1 - T_0 / T$  is the efficiency of an ideal reversible system operating between the effective temperature $T$ of the hot region,  and the constant temperature of the environment, $T_0$. The maximum theoretical efficiency, $\eta_r,$ is known as the Carnot efficiency.    $\Delta \eta$ represents efficiency losses due to irreversibility, a consequence of net entropy generation in the universe, $S_{gen}$.   

From standard thermodynamics theory, the efficiency loss is given by $\Delta \eta = T_0 S_{gen} / Q_{in}$.  Entropy generated in one cycle of operation is 
\[ 
\Delta S_{gen} = \oint \delta Q / T = \int_{in} \delta Q_{in} / T + \int_{out} \delta Q_{out} / T_0 \ = \ \Delta S_{in} - \Delta S_{out} 
\]
In scaling a system such as ours, which supplies input heat via a surface, it is appropriate to hold input power per unit area constant, so that $\dot{Q}_{in} \sim L^{2}$.  For systems that furnish heat via a volume, the appropriate choice is $\dot{Q}_{in} \sim L^3$. 
During the input phase, entropy generation is $\Delta S_{in} = \Delta S_{heat} + \Delta S'_{heat}$, which is a sum of entropy generated by input heating and entropy due to  heat transfer out of the device during the heating phase.  Here, we ignore the latter, as would be appropriate if  heat is delivered over a sufficiently short timescale. Thus, 
\[
\Delta S_{in}  =  \int_{in} \delta Q_{in} / T  = \int_{t_1}^{t_2}  \frac{\dot{Q}_{in} \, dt }{T(t)} \]
For $\dot{Q}_{in}\sim L^2$ or $L^3$, the respective generated entropies are 
\[
\Delta S_{in} \sim L^2 \ \ \ \mathrm{or} \ \ \ \ \Delta S_{in} \sim L^3
\]
Entropy generated in the heat output phase is given by
\[ 
\Delta S_{out} = \int_{out} \frac{\delta Q_{out}}{T_0} = \frac{1}{T_0} \int_{t_2}^{t_1} \, \dot{Q}_{out}(t) \, dt
\]
We consider closed thermodynamic systems for which Fourier's law governs heat transfer output.  If the effective surface area is $A$, one has $\dot{Q}_{out} = - k A \nabla T$.  Thus
\[ 
\Delta S_{out} = \frac{-1}{T_0} \int   k A \nabla T
\]
The temperature gradient and area scale dimensionally as $\nabla T \sim (T - T_0) / L$ and $A\sim L^2,$ so
\[
\Delta S_{out} 
 \sim 
  L^2/L 
   = 
    L \] 
For surface heating, $Q_{in} \sim L^2$ and the total generated entropy  scales as, 
\[ S_{gen}  = S_{out} - S_{in}  \sim  C_1 L + C_2 L^2 \]
The efficiency loss scales as
\[
\Delta \eta = \frac{T_0}{Q_{in}} ( \Delta S_{out} - \Delta S_{in}) \sim  \frac{1}{L^2} (C_1 L + C_2 L^2) \sim L^{-1} + C
\]
For heat delivery via a volume, $Q_{in} \sim L^3$, which yields
\[
 S_{gen}  = S_{in} - S_{out}  \ \sim \  C_1 L + C_2 L^3 
\]
The efficiency loss scales in this case as
\[
\Delta \eta \ \sim \ L^{-2} + C
\]
In each case, the efficiency loss increases as the characteristic dimension $L$ decreases.  This analysis ignores losses arising from sources other than heat transfer, such as viscous losses, or losses such as those noted above.  Nonetheless, efficiency scaling behavior that qualitatively agrees with this analysis  has been observed in many practical thermodynamic systems \cite{sher2009scalinglimitations,burugupally2018actuatorscaling}.

\subsection*{S5. Perceptual experiment design}

During each experiment, stimuli were presented in randomized order. In each trial of every experiment, participants could request that a stimulus be repeated as many times as desired, but did so infrequently. The mean number of stimulus presentations was between 1.13 and 1.36 for all experiments. In experiments 1, 2, 3, 5, and 6 participants viewed graphics on a computer monitor that displayed the possible responses and supplied verbal responses that were logged by the experimenter using a computer terminal. In experiment 4 participants responded using a rotary dial and visual analog scale.  Following the six experiments, participants supplied descriptions of the sensations via free verbalization.  No participant reported feeling heat sensations or reported observing any emitted light. The high response accuracies in experiments 1, 2, 3, 5, and 6 precluded meaningful statistical analysis.  The design, protocol, methods, and analyses used for each experiment were as follows.

\subsubsection*{Experiment 1: Linear Motion Patterns}
The experiment evaluated the perceptual identification of tactile linear motion patterns, along each of four cardinal directions, leftward, rightward, upward or downward, using a four alternative forced choice task.  In each trial, participants reported the direction of motion.  The stimuli were generated through the sequential excitation of three pixels along the indicated direction (Fig.~4A).  Each pixel was excited for 300 ms using cyclic photostimulation with $t_p$ = 20 ms and $t_g$ = 60 ms (see Fig.~3).  The pattern completed in 900 ms, and following a pause of 500 ms, was repeated a second time.  Each stimulus was presented 5 times during the experiment.  Each of the ten participants completed 20 trials, yielding 200 total responses across all participants.  The motion direction was correctly identified in 100\% of trials (chance level: 25\%). 

\subsubsection*{Experiment 2: Perception of Rotational Motion}
The experiment was based on the perceptual discrimination of tactile rotational motion patterns, either clockwise or counterclockwise, at three different motion speeds.  In each trial, participants reported the direction of rotation as clockwise or counterclockwise.  The stimuli were generated through sequential activation of eight pixels around a center pixel, as shown in Fig.~4B, beginning at the bottom left pixel.  The direction of rotation and the motion speed were manipulated for each trial.
Motion speeds of $v =$ 16, 32, and 64 mm/s were realized by exciting each pixel in the sequence for duration $t_a =$ 200, 100, or 50 ms respectively. Each pixel was excited using cyclic photostimulation with $t_p = 25$ ms and $t_g = 71$ ms (see Fig.~3). There were six distinct stimuli.  Each was repeated 5 times, yielding 36 trials per participant, and 360 trials across all participants.  Response accuracy was between 92\% and 98\% for all participants, and mean response accuracy was 94.7\% (341/360 trials correct).  Chance performance was 50\%. 

\subsubsection*{Experiment 3: Spatial Localization}
The experiment was based on the perceptual localization of a single excited pixel located beneath the finger pad of digit 2. In each trial, the center pixel was excited as a reference, and after a pause of 750 ms one of nine target pixels was excited.  Participants reported the location of the target pixel, as a nine alternative forced choice task. Pixels were excited for 500 ms via cyclic photostimulation with $t_p = 25$ ms, and $t_g = 25$ ms (see Fig.~3). There were nine distinct stimuli.  Each was repeated five times, yielding 45 trials per participant, and 450 trials across the experiment.  Response accuracy was between 62\% and 86\% for all participants, and mean response accuracy was 78.4\%  (353/450 trials correct).  Chance performance was 11.1\%.  

\subsubsection*{Experiment 4: Magnitude Estimation of Intensity}
The experiment was based on magnitude estimation of perceived intensity at different optical powers.    In each trial, a pixel at the center of the finger pad of digit 1 was excited.  Participants reported the stimulus intensity using a continuous rotation dial and visual analog scale on a computer monitor.  The scale was labeled to indicate no vibration at one limit and strongest vibration at the opposite limit.  Participants were required to move the dial from its initial position before submitting, and were not permitted to select values at the ends of the scale. The pixel was excited by cyclic photostimulation with $t_p = 25$ ms and $t_g = 25$ ms.  Stimulus duration was 1 second. Power was manipulated for each trial and set to one of nine equally spaced values from $P_L=0.5$ W to $2.5$ W. Each stimulus was repeated 8 times. Each participant completed 72 trials. For each power level $P_L$, a geometric mean was calculated from the response data of each participant. The geometric mean of the participant was then normalized by the grand mean across all power levels to produce a value $I(P_L,k)$ representing perceived intensity at power level $P_L$ for participant $k$ \cite{jones2013psychophysics}. The results were analyzed by statistical regression analysis in order assess the relationship between optical power $P_L$ and intensity $I$. A linear relationship was observed, $I = \alpha P_L + \beta$, with $R^2=0.87$ (Fig.~4D).

\subsubsection*{Experiment 5:  Temporal Pattern Discrimination} 

The experiment was based on temporal pattern discrimination, using a three-interval forced choice task.   In each trial, participants felt a sequence of three stimuli and reported which of the three was the oddball.  Two stimuli in the sequence were identical in temporal pattern (standard stimuli), and one was different (oddball stimulus).  The temporal pattern was produced by cyclic photostimulation with one of two pulse patterns, either slow ($t_p = 35$ ms and $t_g = 35$ ms, $f = 14.3$ Hz) or fast ($t_p = 10$ ms, $t_g = 10$ ms, $f = 50$ Hz). Duration was 400 ms for all stimuli, standard and oddball.   Stimulus order was random for each trial, and the standard and oddball parameters were counterbalanced in the stimulus set.    
Each was presented 4 times, yielding 24 trials per participant and 240 across the experiment. Mean correct response rate across participants was 89.6\% (215/240 correct trials). Correct response rate was between 50\% and 100\% for all participants.   Chance performance was 33.3\%. 

\subsubsection*{Experiment 6: Multi-point Tactile Pattern Perception} 
The experiment was based on the perception of multi-point tactile patterns felt by multiple digits.  In each trial, tactile patterns were presented simultaneously beneath two of the four digits (2 through 5) contacting the display, after which participants reported the pair of locations at which the patterns were presented, by identifying the relevant digits.  The tactile patterns were composed of rapid excitation sequences of four pixels located in a square region beneath each stimulated digit (Fig.~S17). Each pixel excitation was a pulse of duration $t_p = 15$ ms with power $P_L=2.5$ W.  Total stimulus duration was 600 ms. The sequences of excitation at each digit were interleaved in time. There were six distinct stimuli.  Each was presented 5 times, yielding 30 trials per participant and 300 across the experiment.  Mean correct response rate across participants was 96.3\%.  Correct response rate was between 83.3\% and 100\% for all participants.  Chance performance was 16.7\%.

\clearpage

\section*{Supplementary figures}

\begin{figure*}[h]
    \centering
    \includegraphics[width = \linewidth/2]{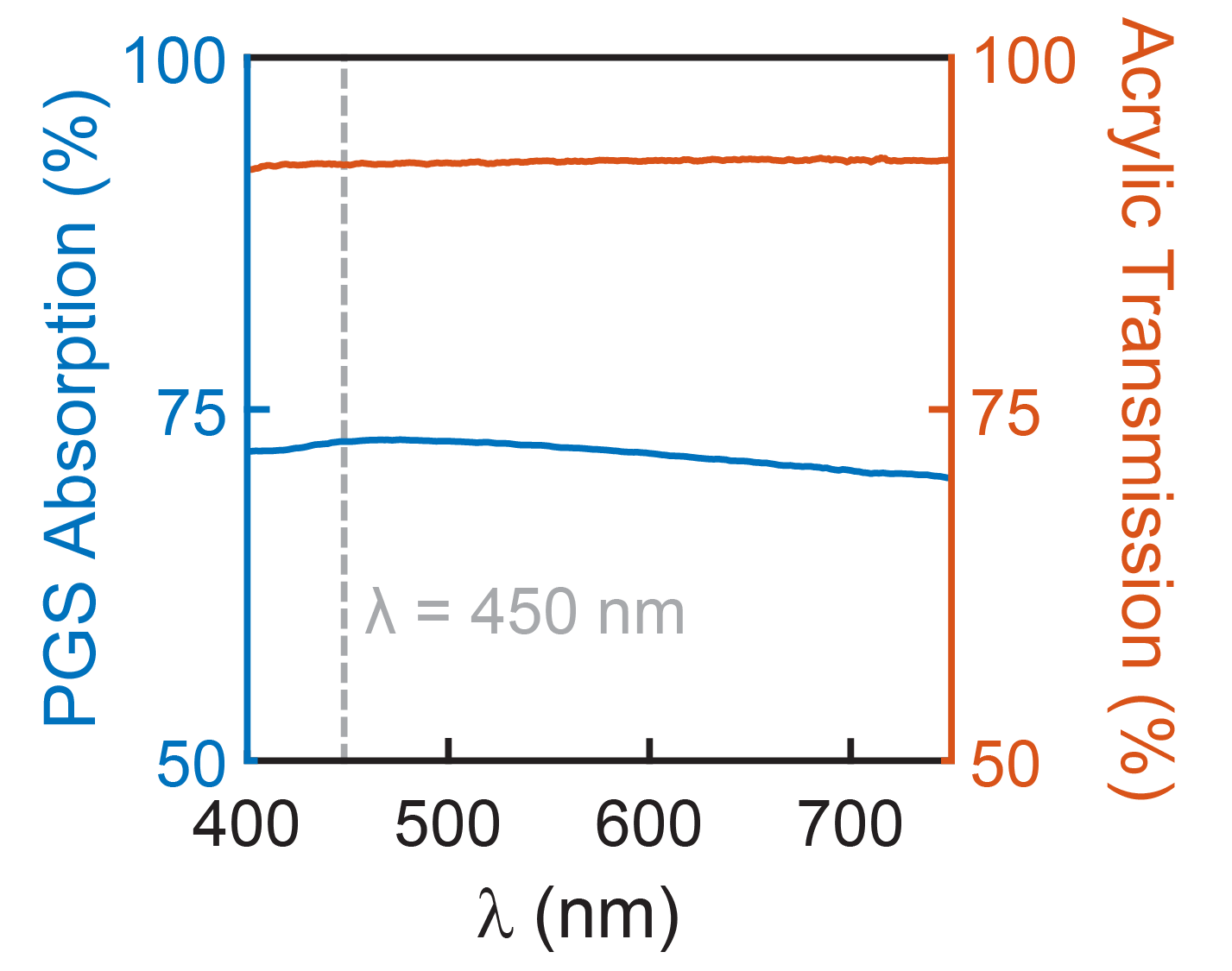}
    \caption{\textbf{Fig. S2. Spectroscopy measurements.} Blue line: PGS Optical absorption spectrum vs. wavelength. Orange line: Acrylic transmission spectrum vs. wavelength. The vertical gray line indicates the wavelength of the laser used to activate a pixel. The total optical absorbance at wavelength $\lambda = 450$ nm is $\varepsilon = 66\%$.}
    \label{supp:spectroscopy}
\end{figure*}

\clearpage
\begin{figure*}
    \centering
    \includegraphics[width = \linewidth/2]{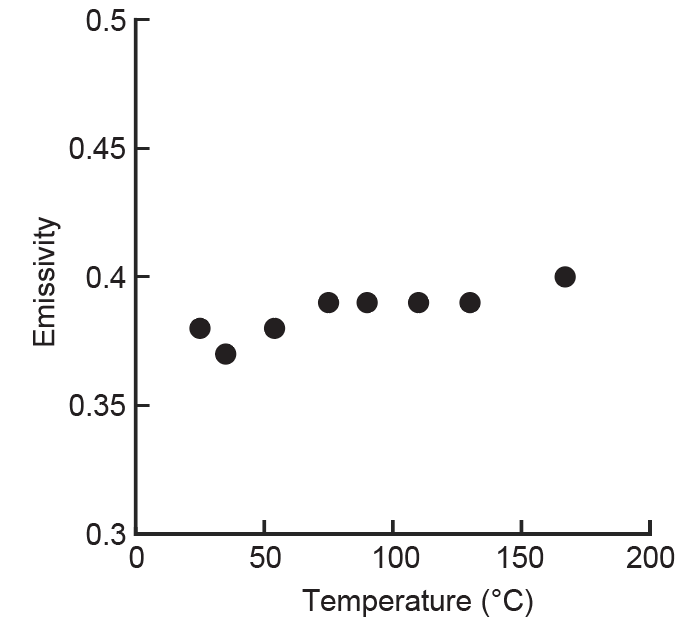}
    \caption{\textbf{Fig. S3. Measured thermal emissivity of PGS.} To determine the emissivity, $\epsilon$, of PGS, we used a thermal control stage (HCP621G, Instec, USA), a MWIR camera (Fast M3K, Telops, USA), and a microscope lens (1x, Telops, USA). We varied the temperature of the control stage and adjusted the emissivity until the measured sample temperature corresponded to the control stage. The emissivity was determined to be 0.39 in the temperature range 25 $^{\circ}$C to 167 $^{\circ}$C.}
    \label{supp:emissivity}
\end{figure*}
\clearpage

\begin{figure*}
    \centering
    \includegraphics[width = \linewidth]{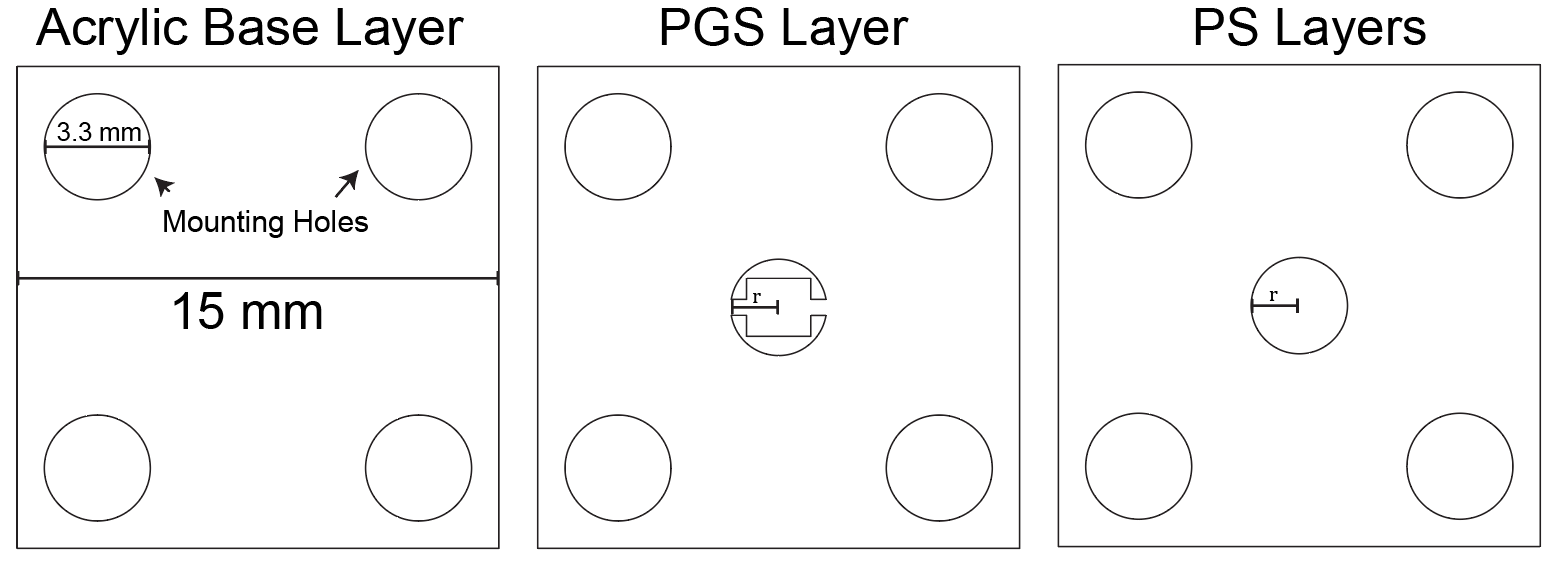}
    \caption{\textbf{Fig. S4. Layered assembly for a 1-pixel display.} The acrylic and PS layers were laser cut, and the absorber layer was patterned using a vinyl cutter.}
    \label{supp:single-sample-layers}
\end{figure*}

\clearpage
\begin{figure*}
    \centering
    \includegraphics[width = \linewidth]{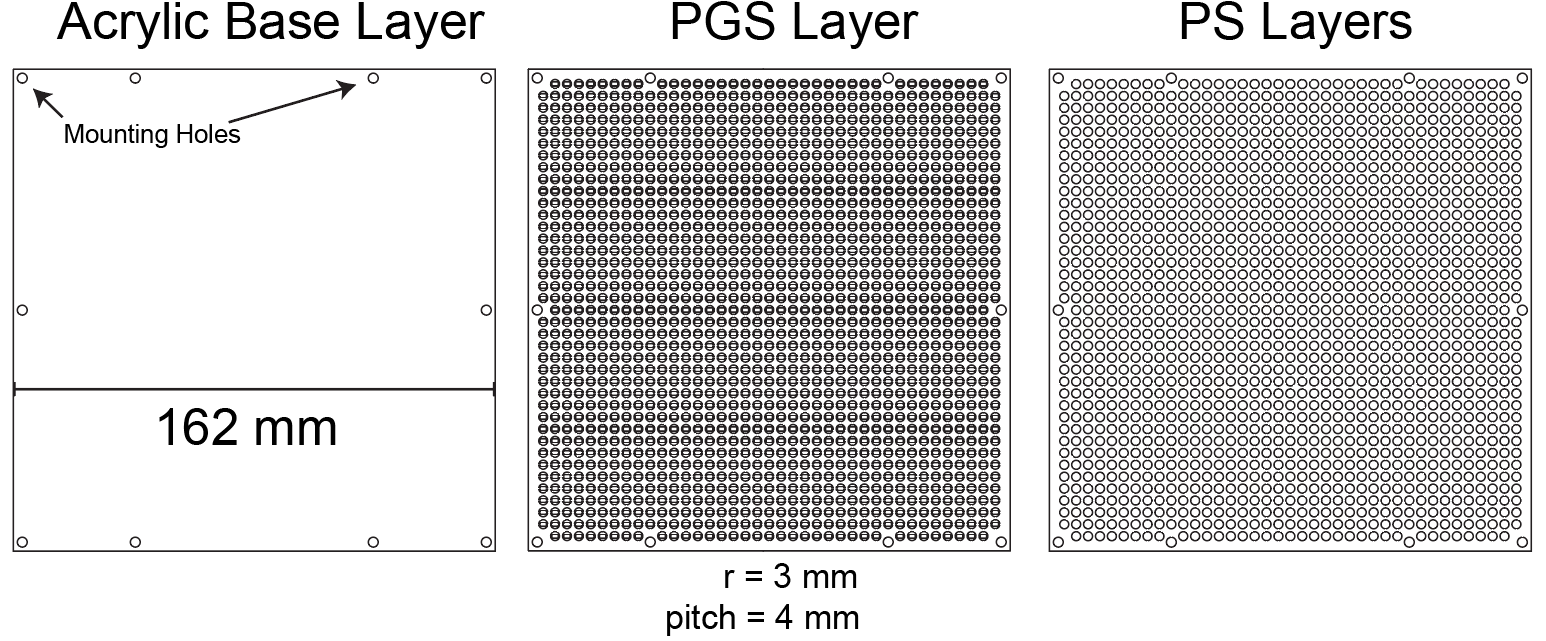}
    \caption{\textbf{Fig. S5. Layer designs of the 1,511-pixel display.} A square grid of 39 x 39 holes was laser-cut into both the top and bottom PS sheets. The PGS was patterned with a vinyl cutter to create a matching lattice of photoabsorbers, aligning them with the circular features on the silicone sheets.}
    \label{supp:array-sample-layers}
\end{figure*}

\clearpage
\begin{figure*}
    \centering
    \includegraphics[width = \linewidth]{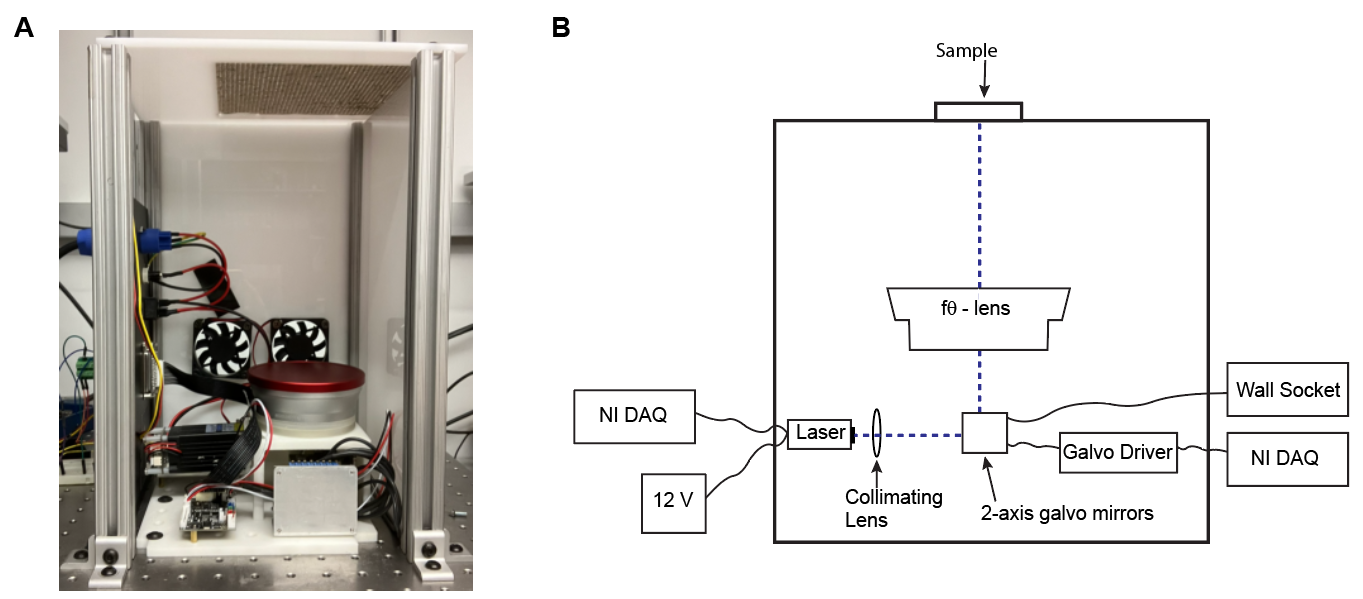}
    \caption{\textbf{Fig. S6. Internal components and optical path in tactile display} A) Photo of the optical stage. The f-theta lens has a red lens-cover on. B) Schematic of the optical stage, not to scale.}
    \label{supp:setup}
\end{figure*}

\clearpage
\begin{figure*}
    \centering
    \includegraphics[width = \linewidth]{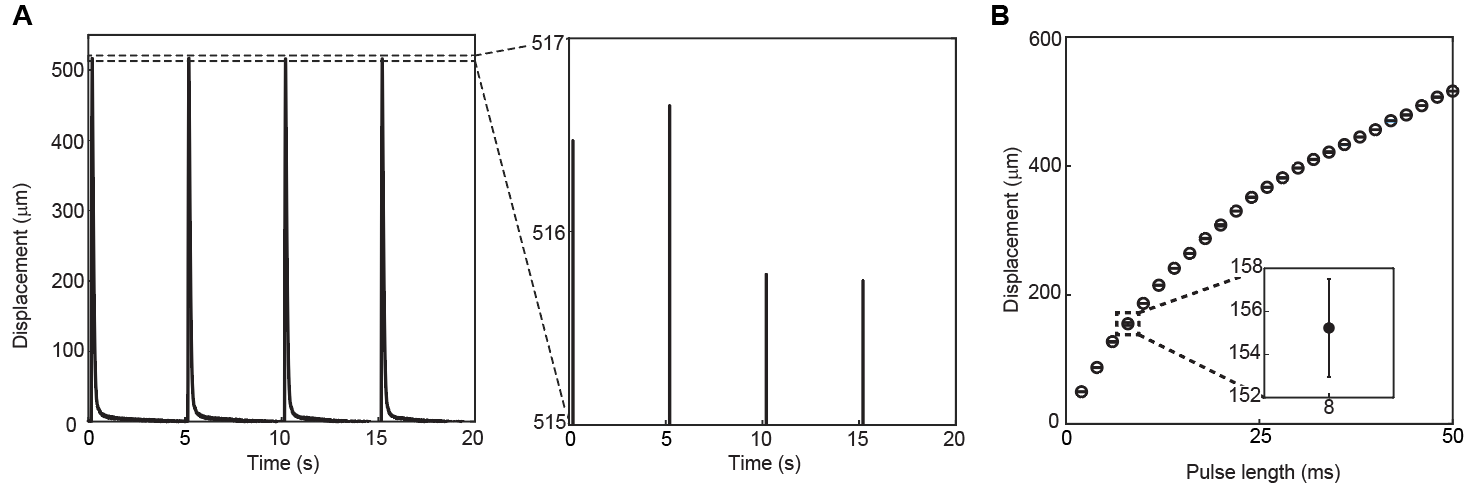}
    \caption{\textbf{Fig. S7. Photomechanical response repeatability across cycles.} A) Displacement under cyclic photostimulation (4 pulses, $t_p = 50$ ms, $t_g = 4.95$ s, $w = 0.4$ mm). Enlarged area reveals modest variation in amplitude of the four peaks. Amplitude difference: $<$ 0.5 $\mu$m. B) Peak displacement ($n = 4$) vs.~pulse length, $t_p$, ranging from 4 ms to 50 ms. Error bars: 1 standard deviation. Inset: $t_p = 8$ ms, which has the largest fractional uncertainty ($d = 155.2 \pm 2.3$ $\mu$m).}
    \label{supp:uncertainty}
\end{figure*}

\clearpage
\begin{figure*}
    \centering
    \includegraphics[width = \linewidth]{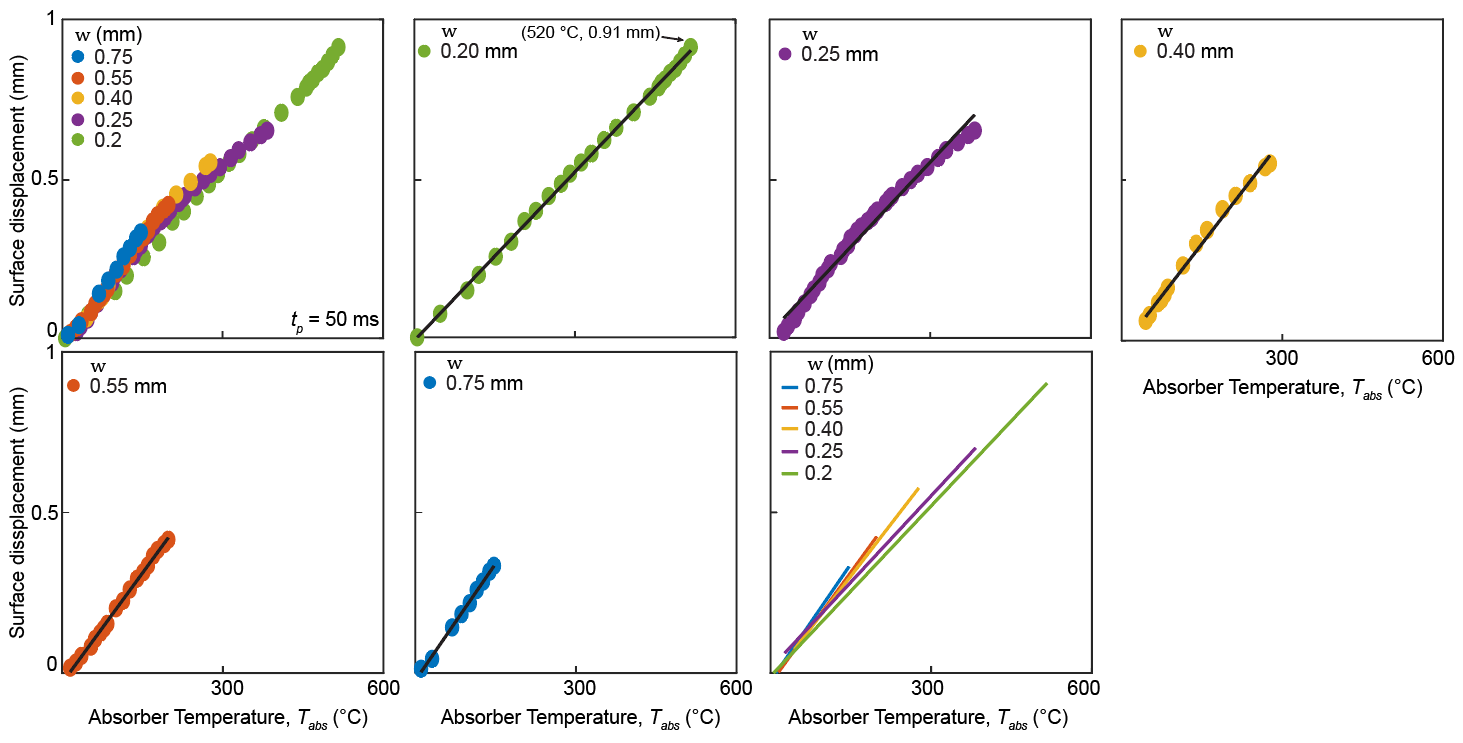}
    \caption{\textbf{Fig. S8. Experimentally observed  relationship between absorber temperature and membrane displacement.} The optotactile pixels exhibit a nearly static relationship between absorber temperature and membrane displacement across a range of temperatures and geometries.}
    \label{supp:linearity}
\end{figure*}

\clearpage
\begin{figure*}
    \centering
    \includegraphics[width = 0.5\linewidth]{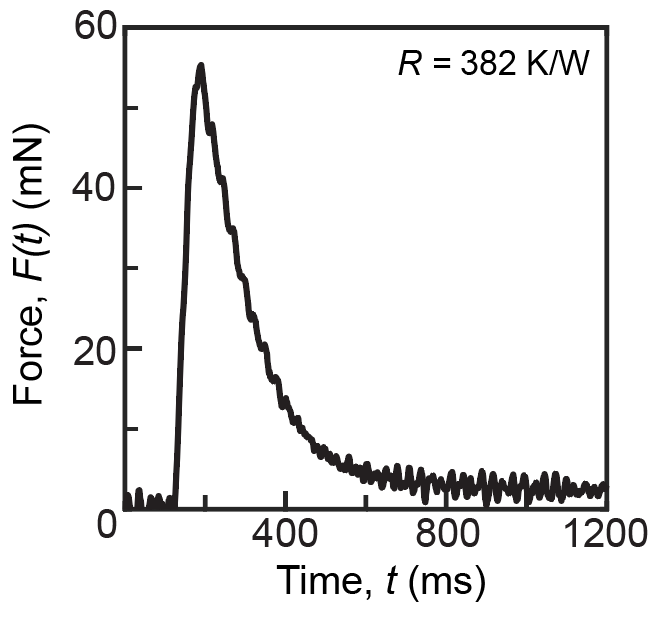}
    \caption{\textbf{Fig. S9. Isometric force $\bm{R = 382}$ K/W ($\bm{w = 0.2}$ mm).} Photostimulation of a higher-resistance pixel yielded a peak force of $F=55$ mN after light stimulation for $50$ ms.}
    \label{supp:force_w02}
\end{figure*}

\clearpage
\begin{figure*}
    \centering
    \includegraphics[width = \linewidth]{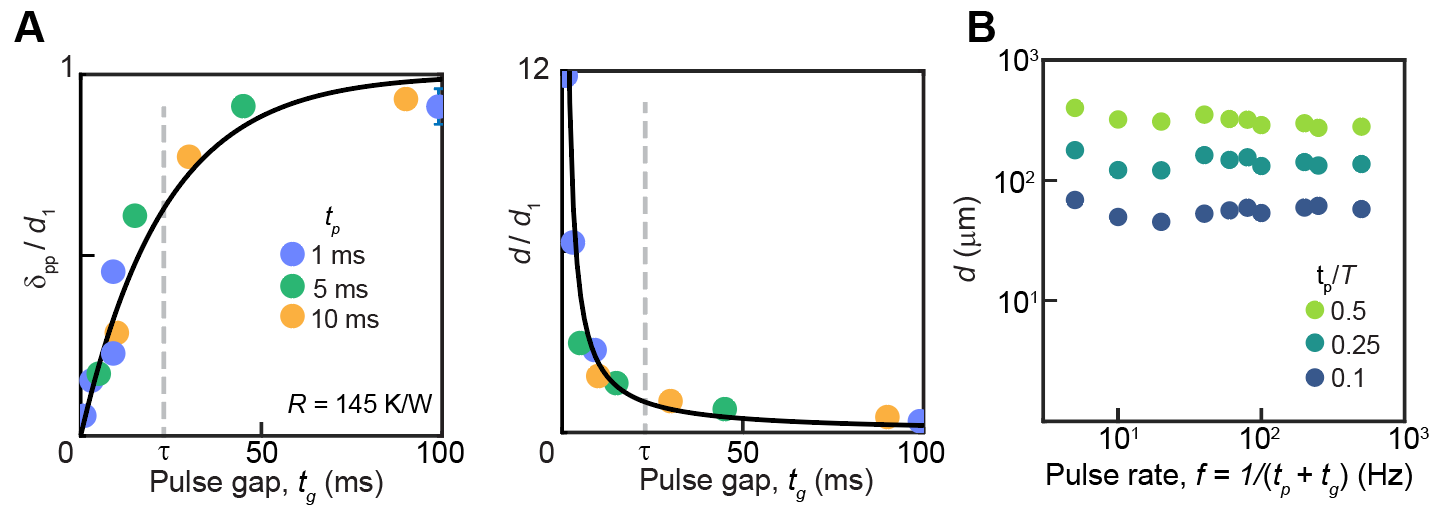}
    \caption{\textbf{Fig. S10. Slow and oscillating components of membrane deflection under cyclic photostimulation} A) The relative magnitude of slow and oscillatory displacement can be adjusted by controlling the pulse gap, $t_g$, and is independent of pulse length when normalized by $d_1$. B) Slow component amplitude after 100 pulses is determined by duty cycle, $\nu$, since the time-averaged added energy is invariant to pulse rate, $f$.}
    \label{supp:peak-to-peak}
\end{figure*}

\clearpage
\begin{figure*}
    \centering
    \includegraphics[width = 0.5\linewidth]{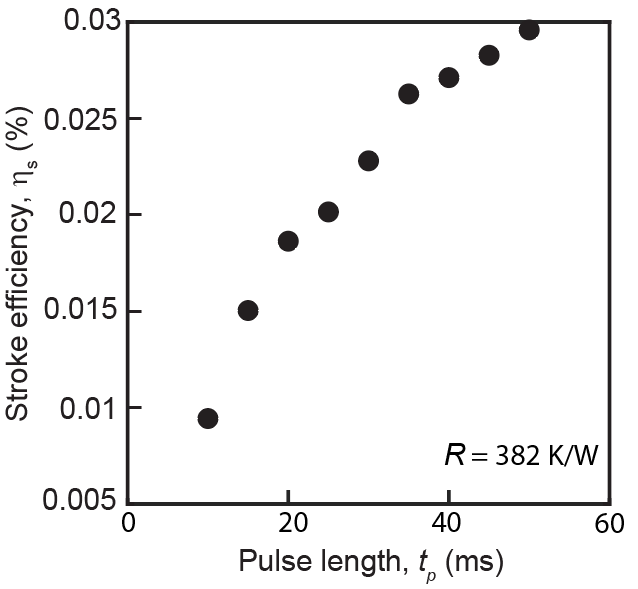}
    \caption{\textbf{Fig. S11. S efficiency as a function of pulse length, $t_p$.} Stroke efficiency increases sublinearly with pulse duration.}
    \label{supp:efficiency}
\end{figure*}

\clearpage
\begin{figure*}
    \centering
    \includegraphics[width = \linewidth]{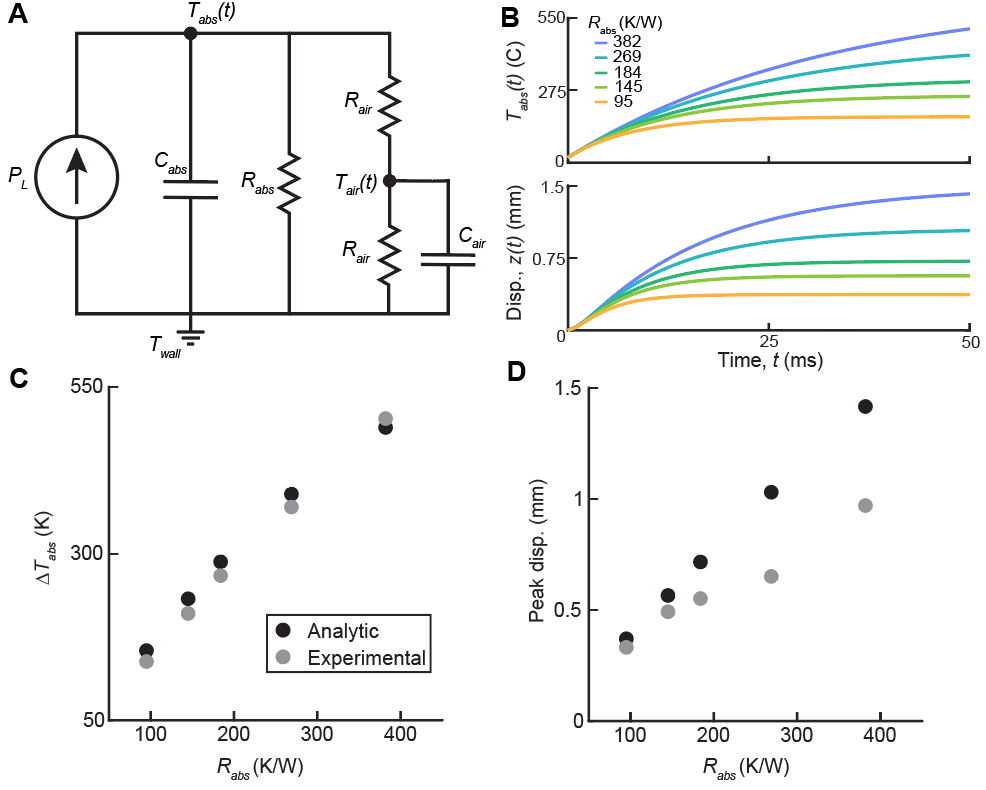}
    \caption{\textbf{Fig. S12. Effective heat transfer description and comparison with experiments.} A) Equivalent circuit thermal network. B) Photoabsorber temperature and membrane displacement vs.~time.  Heat pulse duration $t_p=50$ ms. C,D) Comparison with experimental findings.}
    \label{supp:analytic model}
\end{figure*}

\clearpage
\begin{figure*}
    \centering
    \includegraphics[width = 0.5\linewidth]{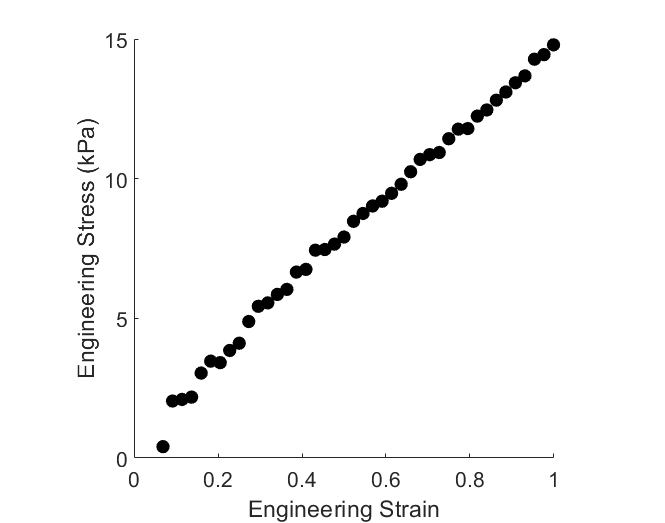}
    \caption{\textbf{Fig. S13. Linear elasticity of EcoFlex 00-10.}  Adapted from Marechal et al. \cite{marechal2021}.  The range of stresses and engineering strains encompasses those arising in our device.}
    \label{supp:ecoflex-linearity}
\end{figure*}
\clearpage
\begin{figure*}
    \centering
    \includegraphics[width = \linewidth]{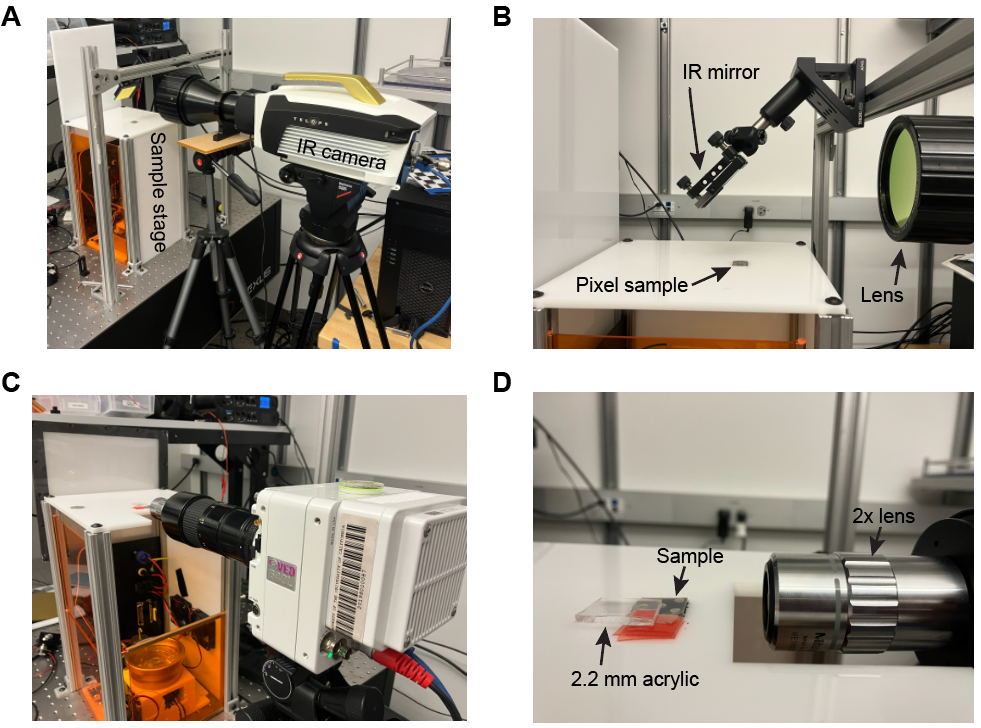}
    \caption{\textbf{Fig. S14. Imaging setups.} A,B) Thermal imaging setup. C,D) High-speed video setup.}
    \label{supp:imaging-setup}
\end{figure*}

\clearpage
\begin{figure*}
    \centering
    \includegraphics[width = \linewidth]{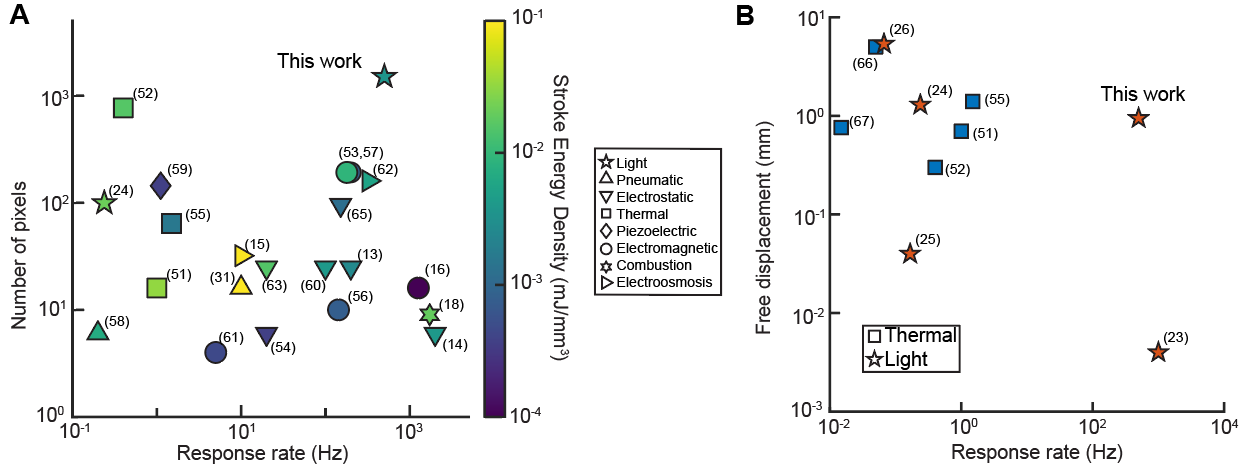}
    \caption{\textbf{Fig. S15. Comparison with other devices}  A) Comparison with prior tactile display systems. Our display can achieve refresh rates and pixel counts that compare favorably to prior work. Corresponding data can be found in Table S1. B) Each optotactile pixel in our display exhibits rapid response, comparing favorably to prior thermal tactile actuation methods. Corresponding data can be found in Table S2. (Star at lower-right: No perceptual results reported. Tactile feedback unlikely to be felt during use, because the reported displacements are smaller than vibration noise in buildings within ISO standard 2631-2:2003 \cite{ISO13586}.)}
    \label{supp:Comparison}
\end{figure*}

\clearpage
\begin{figure*}
    \centering
    \includegraphics[width = 0.75\linewidth]{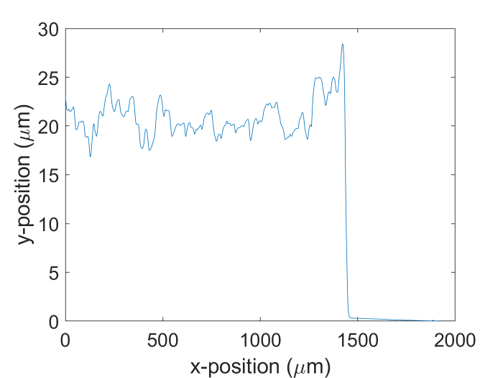}
    \caption{\textbf{Fig. S16. Thickness of PGS with adhesive.} The surface profile of one PGS sample, with adhesive, as measured via profilometry. Each sample was placed on a glass slide with the adhesive side down in the profilometer. The surface profile was collected, starting $\sim 1.5$ mm from the sample edge, and measured for 2 mm. The sudden drop in y-position around $x = 1500 \mu$m represents the edge of the PGS sample. This data was used to characterize the PGS thickness and density as reported in Methods.}  \label{supp:profilometer}
\end{figure*}

\clearpage
\begin{figure*}
    \centering
    \includegraphics[width = 0.65\linewidth]{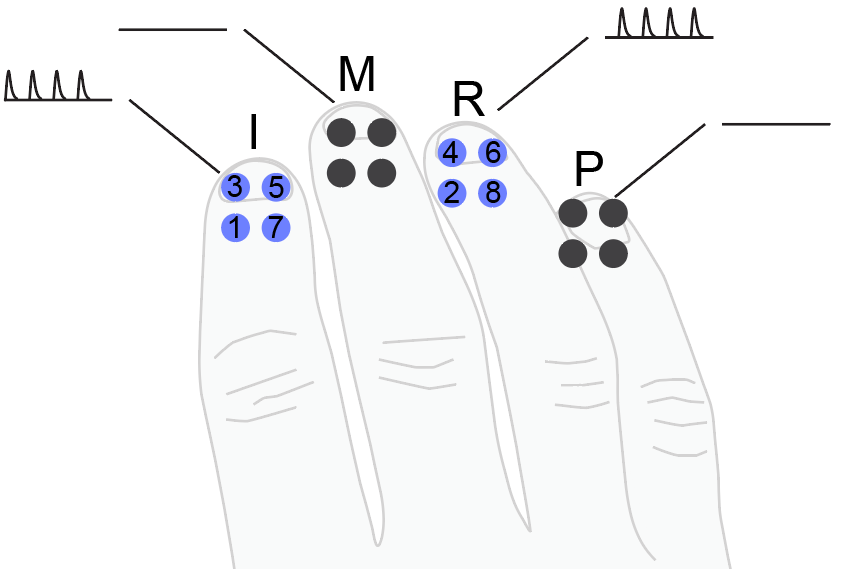}
    \caption{\textbf{Fig. S17. Experiment 6 stimulus.} The stimulus in experiment 6 of the perceptual study was created by sequential activation of the pixels highlighted in blue, with the numbers determining the order of activation. The pulse length was $t_p = 15$ ms. The sequential activation of the pixels was repeated 5 times during the full stimulus.}
    \label{supp:stim_order}
\end{figure*}

\clearpage
\begin{figure*}
    \centering
    \includegraphics[width = \linewidth]{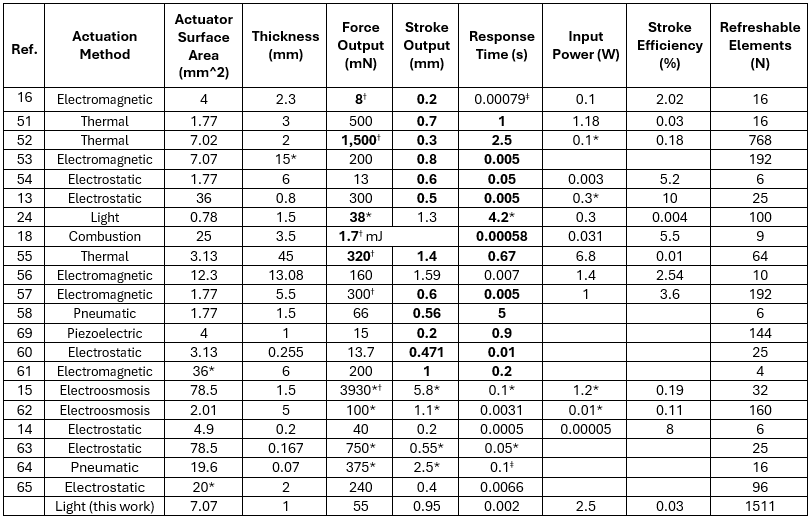}
    \caption{\textbf{Table S1. Comparison table between tactile display systems.} The dataset used to generate the performance comparison between tactile displays (Figs. 1, S15A) \cite{streque2012emactuator,qiu2018polymerdisplay,besse2017fea,zarate2017taxels,han2020dielectric,leroy2020haxel,torras2014optomechanical,heisser2021combustion,velazquez2005shapememory,strasnick2016magneticactuators,kim2020emactuator,wu2012braille,kato2007polymeractuator,koo2008soft,gallo2015display,schultz2023flatpanel,shen2023fluidreality,grasso2023haxel,leroy2023haxel,wang2024metaverse,phung2017actuator}. Bold quantities are derived from the same measurement and are experimentally correlated. Pictorially, graphically, or quantitatively inferred quantities are denoted with (*). Non-blocked force measurements are denoted with (†). Response times which are the reported maximum or resonant frequencies are denoted with (‡). Stroke efficiency is calculated from given quantities. Empty cells reflect values that could not be found. This dataset has been adapted from \cite{heisser2021combustion}.}
    \label{supp:table1}
\end{figure*}

\clearpage
\begin{figure*}
    \centering
    \includegraphics[width = \linewidth]{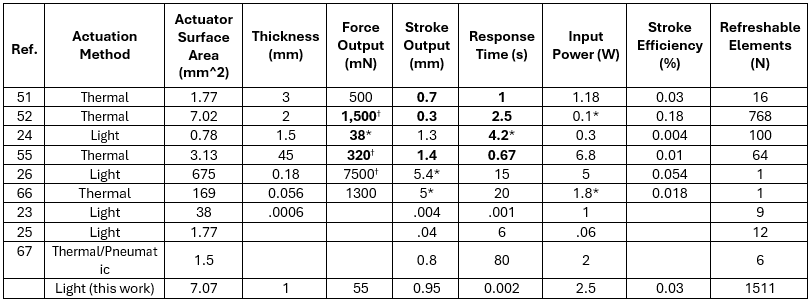}
    \caption{\textbf{Table S2. Comparison table between thermally operating actuators.} The dataset used to generate the performance comparison between thermal actuators (Fig. S15B) is shown here \cite{qiu2018polymerdisplay,besse2017fea,torras2014optomechanical,velazquez2005shapememory,hiraki2020laserpouch,uramune2022hapouch,hwang2021bimorph,camargo2012opticalbraille,soule2016phasebraille}. Bold quantities are derived from the same measurement and are experimentally correlated. Pictorially, graphically, or quantitatively inferred quantities are denoted with (*). Non-blocked force measurements are denoted with (†). Response times which are the reported maximum or resonant frequencies are denoted with (‡). 
    Stroke efficiency is calculated from given quantities. Empty cells reflect values that could not be found. This dataset has been adapted from \cite{heisser2021combustion}.}
    \label{supp:table2}
\end{figure*}

\clearpage
\begin{table}[]
\begin{center}
\begin{tabular}{|c|c|c|c|}
\hline
\textbf{$w\pm 0.05$ (mm)} & $R$ (K/W) & $C$ ($\mu$J/K) & $\tau=RC (ms)$\\ \hline
0.20 & 382 & 81 & 31 \\ \hline
0.25 & 269 & 106 & 29 \\ \hline
0.40 & 184 & 126 & 23 \\ \hline
0.55 & 145 & 99 & 14 \\ \hline
0.75 & 95 & 102 & 10 \\ \hline
\end{tabular}
\caption{\textbf{Table S3. Experimentally determined thermal parameters for photoabsorbers with different bridge widths.} As bridge width, $w,$ increases, both resistance $R$ and time constant, $\tau,$ decrease. Heat capacity values varied from sample to sample, with $C = 101 \pm 16$ $\mu$ JK$^{-1}$ across samples. Parameters were determined using measured data, as described in Text S2. 
}
 
\end{center}
\end{table}

\clearpage
\begin{table}[]
\begin{center}
\begin{tabular}{|c|c|c|c|c|c|c|}
\hline
\textbf{Material} & $c_p$ (J/kg$\cdot$W) & $\rho$ (kg/m$^3$) & $k$ (W/m$\cdot$K) & $E$ (kPa) & $\nu$& Ref. \\ \hline
PGS & 850 & 1962 & See methods & N/A&N/A & \cite{Panasonic_PGS_2024}\\ \hline
EcoFlex 00-10 & 1000 & 1060 & 29 &50&0.49 & \cite{vaicekauskaite2020ecoflex}\\ \hline
PS & 1460 & 970 & 0.092 & 750&0.49 & \cite{vishnuchandar2020polysiloxane}\\ \hline
Acrylic & 1470 & 1190 & 0.18 & N/A & N/A & \cite{comsol61} \\ \hline
\end{tabular}
\caption{\textbf{Table S4. Material parameters used in numerical experiments.} Notation: Specific heat capacity, $c_p$; density $\rho$; thermal conductivity $k$; Young's modulus $E$;  Poisson ratio $\nu$. Values obtained from indicated sources where noted; See Methods for $k_{pgs}$ details.}  
 
\end{center}
\end{table}

\begin{table}[]
\begin{center}
\begin{tabular}{|c|c|c|c|c|}
\hline
\textbf{Material} & \textbf{Pricing} & \textbf{Quantity} &  \textbf{Cost} &  \textbf{Manufacturer} \\ \hline
PGS & $\sim$\$0.16 / cm$^2$  & 262 cm$^2$& \$42& Panasonic \\ \hline
EcoFlex 00-10 & $\sim$\$0.04 / g  & 6.5 g & \$0.26 & Smooth-On\\ \hline
PS & $\sim$\$0.04 / cm$^2$  & 262 cm$^2$& \$10.5& McMaster-Carr (supplier)\\ \hline
Acrylic & $\sim$\$0.01 / cm$^2$  & 262 cm$^2$& \$2.62 & McMaster-Carr (supplier)\\ \hline
\end{tabular}
\caption{\textbf{Table S5. Display surface material costs.} Representative material costs for the surface assembly of the 1,511-pixel tactile display.}  
\end{center}
\end{table}

\begin{table}[]
\begin{center}
\begin{tabular}{|c|c|}
\hline
\textbf{Material} & \textbf{Cost} \\ \hline
Diode laser, 450 nm, 4W & \$45 \\ \hline
F$\Theta$ lens& \$289\\ \hline
Collimating lens & \$8\\ \hline
Scanning galvo system \& Drive electronics & \$75 \\ \hline
\end{tabular}
\caption{\textbf{Table S6. Optical stage costs.} Representative component costs for the optical stage used to drive the demonstrated displays.} 
\end{center}
\end{table}

\newpage
\clearpage

\begin{figure*}
    \centering
    \caption{\textbf{Movie S1. Thermo-Mechanical Response of Pixel Under Photostimulation.} Left: Time-resolved photoabsorber temperature, captured via thermal imaging.  Right: Imaged embrane deflection. The photoabsorber bridge width is $w = 0.75$mm, Optical pulse duration $t_p = 50$ms. The right video shows the corresponding membrane deflection. Scale bar: 1 mm.  Video playback speed: 1/75x.}
    \label{supp:Movie1}
\end{figure*}

\newpage
\begin{figure*}
    \centering
    \caption{\textbf{Movie S2. High-speed video of membrane displacement for different photoabsorber geometries.} Bridge widths: $w = 0.2, 0.25, 0.4, 0.55, 0.75$ mm. Optical pulse duration: $t_p = 50$ ms. Video playback speed: 1/50x.}
    \label{supp:Movie2}
\end{figure*}

\newpage
\begin{figure*}
    \centering
    \caption{\textbf{Movie S3. High-speed video of mm-scale photomechanical response of optotactile pixel.} The maximum displacement of 0.97 mm approximately coincides with cessation of photostimulation. Bridge width: $w = 0.2$ mm. Optical pulse duration: $t_p = 50$ ms. Video playback speed: 1/50x.}
    \label{supp:Movie3}
\end{figure*}

\clearpage
\newpage


\end{document}